\input jytex.tex   
\typesize=10pt
\magnification=1200
\baselineskip17truept
\footnotenumstyle{arabic}
\hsize=6truein\vsize=8.5truein
\sectionnumstyle{blank}
\chapternumstyle{blank}
\chapternum=1
\sectionnum=1
\pagenum=0

\def\begintitle{\pagenumstyle{blank}\parindent=0pt\begin{narrow}[0.4in]}
\def\endtitle{\end{narrow}\newpage\pagenumstyle{arabic}}


\def\beginexercise{\vskip 20truept\parindent=0pt\begin{narrow}[10
truept]}
\def\endexercise{\vskip 10truept\end{narrow}}


\def\eql#1{\eqno\eqnlabel{#1}}
\def\ref{\reference}
\def\peq{\puteqn}
\def\pref{\putref}

\def\mgn{\marginnote}
\def\bex{\begin{exercise}}
\def\eex{\end{exercise}}


\font\open=msbm10 
\font\ssb=cmss10
\def\mbox#1{{\leavevmode\hbox{#1}}}

\def\hspace#1{{\phantom{\mbox#1}}}
\def\oR{\mbox{\open\char82}}

\def\oZ{\mbox{\open\char90}}

\def\ssse{\mbox{{\ssb\char101}}}

\def\al{\alpha}
\def\be{\beta}
\def\ga{\gamma}

\def\Ga{\Gamma}

\def\ep{\epsilon}

\def\ka{\kappa}
\def\la{\lambda}

\def\om{\omega}
\def\Om{\Omega}

\def\th{\theta}

\def\ze{\zeta}

\def\De{\Delta}

\def\Real{{\rm Re\,}}

\def\zs{{\rm zs }}
\def\dc{{\rm dc }}
\def\dn{{\rm dn }}
\def\cs{{\rm cs }}
\def\sn{{\rm sn }}

\def\cn{{\rm cn }}
\def\ns{{\rm ns }}
\def\ds{{\rm ds }}

\def\Imag{{\rm Im\,}}

\def\zf{$\zeta$--function}
\def\zfs{$\zeta$--functions}


\def\frac#1/#2{\leavevmode\kern.1em
\raise.5ex\hbox{\the\scriptfont0 #1}\kern-.1em/\kern-.15em
\lower.25ex\hbox{\the\scriptfont0 #2}}
\def\sfrac#1/#2{\leavevmode\kern.1em
\raise.5ex\hbox{\the\scriptscriptfont0 #1}\kern-.1em/\kern-.15em
\lower.25ex\hbox{\the\scriptscriptfont0 #2}}

\def\gtorder{\mathrel{\raise.3ex\hbox{$>$}\mkern-14mu
             \lower0.6ex\hbox{$\sim$}}}
\def\ltorder{\mathrel{\raise.3ex\hbox{$<$}\mkern-14mu
             \lower0.6ex\hbox{$\sim$}}}

\def\semidirprod{\rlap{\ss C}\raise1pt\hbox{$\mkern.75mu\times$}}
\def\for{\lower6pt\hbox{$\Big|$}}
\def\fish{\kern-.25em{\phantom{abcde}\over \phantom{abcde}}\kern-.25em}


\def\boxit#1{\vbox{\hrule\hbox{\vrule\kern3pt
        \vbox{\kern3pt#1\kern3pt}\kern3pt\vrule}\hrule}}
\def\dalemb#1#2{{\vbox{\hrule height .#2pt
        \hbox{\vrule width.#2pt height#1pt \kern#1pt
                \vrule width.#2pt}
        \hrule height.#2pt}}}

\def\frac#1#2{{{#1}\over{#2}}}

\def\comb#1#2{{\left(#1\atop#2\right)}}

\def\etc{{\it etc. }}

\def\eg{{\it e.g. }}
\def\ie{{\it i.e. }}
\def\cf{{\it cf }}
\def\pa{\partial}


  %

\def\sumdasht#1#2{{\mathop{{\sum}'}_{#1}^{#2}}}

\def\3j#1#2#3#4#5#6{\left\lgroup\matrix{#1&#2&#3\cr#4&#5&#6\cr}
\right\rgroup}

\def\man{{\cal M}}

\def\m?{\mgn{?}}

\def\pa{\partial}

\def\beq{\begin{eqnarray}}
\def\eeq{\end{eqnarray}}


\def\cqg#1#2#3{{\it Class. Quant. Grav.} {\bf {#1}} ({#2}) #3}

\def\jmp#1#2#3{{\it J. Math. Phys.} {\bf {#1}} ({#2}) #3}
\def\jpa#1#2#3{{\it J. Phys.} {\bf A{#1}} ({#2}) #3}

\def\np#1#2#3{{\it Nucl. Phys.} {\bf B{#1}} ({#2}) #3}
\def\pl#1#2#3{{\it Phys. Lett.} {\bf {#1}} ({#2}) #3}

\def\prp#1#2#3{{\it Phys. Rep.} {\bf {#1}} ({#2}) #3}
\def\pr#1#2#3{{\it Phys. Rev.} {\bf {#1}} ({#2}) #3}

\def\prD#1#2#3{{\it Phys. Rev.} {\bf D{#1}} ({#2}) #3}

\def\prs#1#2#3{{\it Proc. Roy. Soc.} {\bf A{#1}} ({#2}) #3}

\def\aim#1#2#3{{\it Adv. in Math.} {\bf {#1}} ({#2}) #3}

\def\jram#1#2#3{{\it J. f. reine u. Angew. Math.} {\bf {#1}} ({#2}) #3}
\def\jims#1#2#3{{\it J. Indian. Math. Soc.} {\bf {#1}} ({#2}) #3}
\def\jlms#1#2#3{{\it J. Lond. Math. Soc.} {\bf {#1}} ({#2}) #3}

\def\ma#1#2#3{{\it Math. Ann.} {\bf {#1}} ({#2}) #3}

\def\plms#1#2#3{{\it Proc. Lond. Math. Soc.} {\bf {#1}} ({#2}) #3}

\def\rmjm#1#2#3{{\it Rocky Mountain J. Math.} {\bf {#1}} ({#2}) #3}

\begin{title}
\vglue 1truein
\vskip15truept
\centertext{\Bigfonts \bf Elliptic functions and temperature}\vskip10truept
\centertext{\Bigfonts \bf inversion symmetry on spheres}

 \vskip 20truept
\centertext{J.S.Dowker\footnote{dowker@a35.ph.man.ac.uk} }
\vskip 7truept \centertext{\it Department of Theoretical Physics, }
\centertext{\it The University of Manchester,} 
\centertext{ \it Manchester, England} \vskip 10truept \centertext{and}
\vskip7truept
\centertext{Klaus Kirsten\footnote{kirsten@mis.mpg.de}}
\vskip7truept\centertext{\it Max Planck Institute for 
Mathematics in the Sciences, }
\centertext{\it Inselstrasse 22-26, 04103 Leipzig, Germany}
 \vskip15truept
\begin{narrow}
Finite temperature boson and fermion field theories on the space-time
manifolds $\oR\times$S$^d$ are discussed with one eye on the questions of
temperature inversion symmetry and modular invariance. For conformally
invariant theories it is shown that the total energy at any temperature for
any dimension, $d$, is given as a power series in the $d=3$ and $d=5$
energies, for scalars, and the $d=1$ and $d=3$ energies for spinors.
Further, these energies can be given in finite terms at specific
temperatures associated with singular moduli of elliptic function theory.
Some examples are listed and numbers given.
\end{narrow}
\vskip 5truept
\vskip 60truept
\vfil
\end{title}
\pagenum=0
\section{\bf1. Introduction.}

The symmetry between low and high temperatures in finite temperature field
theory has re-emerged recently in connection with the questions of entropy
bounds and the Verlinde--Cardy relation. Some comments were made in an
earlier paper [\pref{Dow4}] which we wish to enlarge on here. Although
the main point of this earlier communication was the thermodynamic
significance of any zero modes, this will occupy us only incidentally here.

The primary emphasis in the present article is on inversion symmetries in
finite temperature field theory and its relation to elliptic functions.
Inversion symmetry usually appears as a consequence of Jacobi's inversion
identity for theta functions, in one of its many guises, and it should be no
surprise that elliptic functions feature in our discussion.

In Section 2 we discuss for which manifolds simple inversion properties
might be expected and we are naturally led to the manifold S$^1 \times$
S$^d$, S$^1$ being the thermal circle and S$^d$ the $d$-dimensional sphere.
A unified approach to the energy  of quantum fields in curved space-time is
via the thermal $\zeta$-function and some basics are presented in Section
3. Afterwards we relate the energy to Eisenstein series which allow
symmetry properties to be discussed very easily. In particular, we can use
relations between the Eisenstein series and the Weierstrass $\wp$-function
to find the scalar energy for all dimensions just in terms of the results
on the three and five sphere. For spinors one iterates from the one and
three dimensional spheres. Elliptic functions are then used in Section 6
and 7 to give a description of temperature inversion in terms of the
modulus, $k$. The relation to elliptic functions also allows one to express
the energy in terms of powers of $k^2$ and of the complete elliptic
integral, $K(k)$. This has the advantage that, at temperatures
corresponding to singular moduli, closed forms for the energies can be
found, see Section 9. The Conclusion describes the main points raised in
the article. Three appendices contain technicalities about  different
viewpoints on the approaches for spinor fields, power series expansions of
elliptic functions, and some explicit expressions too cumbersome for the
main body of our paper.

\section{\bf 2. Background technical facts.}

Any symmetry relation has to be expressed in a dimensionless quantity say
$aT=a/\be\equiv\xi/2\pi$, where $a$ is some length scale associated with
the spatial manifold, $\man$. Clearly reciprocal symmetry involves
exchanging $\be$ and $a$.

Such a symmetry is therefore suggested whenever one can detect the torus
structure, S$^1\times$S$^1$, in the thermal manifold, S$^1\times \man$, the
above exchange corresponding to a transposition of the two circles. An
example where the symmetries are clearly visible is when $\man$ is the
torus, S$^1\times$S$^1\times \ldots\times$S$^1$. However the case that
interests us here is when $\man$ is the $d$--sphere, in particular $d$ odd.
The sphere, and all spaces of constant curvature, have only one relevant
length scale, which at least simplifies the possibilities for
reciprocation.

A basic analytical fact is that the scalar spectral quantities such as the
heat--kernel (or quantum propagator in our case) on the $d$--sphere can be
`integrated' in steps of two down to that on the circle, for theories
conformal on the space--time. The thermal space, S$^1\times$S$^d$, reduces
to S$^1\times$S$^1$. Alternatively, the thermal circle, S$^1$, can be
differentiated to give a ``thermal $d$--sphere" resulting in the
symmetrical combination S$^d\times$S$^d$. This occurs only if $d$ is odd
and explains why inversion symmetries arise in this case. For even $d$, the
scalar S$^1$ heat-kernel can be fractionally differentiated to give S$^d$,
but only if it is twisted, \ie a $\theta_2$--function, \eg
[\pref{Camporesi}]. This is clearly the origin of Cardy's use of
anti-periodic conditions for bosons, [\pref{Cardy}], for, in order to
achieve any symmetry, the thermal circle must be twisted as well.

A practical strategy is to treat the terms in the degeneracy polynomial
independently and deal with them one by one. Each part will have a
different inversion property. It is this approach that will occupy us
mostly in this paper.

We might mention that these calculations can be extended to {\it
split-rank} symmetric spaces of which odd spheres are the simplest
examples. (See, in this connection, [\pref{Camporesi}].) The heat-kernel
expansion terminates at the {\it first} term on all (semi)-simple group
manifolds, but only if the scalar is conformal in 4 dimensions, whatever
the dimension of the group.
\section{\bf 3. Statistical mechanics.}

The essential points were made in [\pref{Dow4}] but have to be repeated
here for completeness and ease of exposition. A unified approach is via the
thermal \zf, $\ze(s,\be)$, [\pref{DandKe,Gibbons,GandP}], which neatly
incorporates any ultra-violet and zero mode effects.

The general form of $\ze(s,\be)$ is, on an ultrastatic space-time,
T$\times\man$,
  $$
  \ze(s,\be)={i\over\be}\sumdasht{{m=-\infty\atop n}}
{\infty}{d_n\over
 \big(\om_n^2+4\pi^2m^2/\be^2\big)^s}\,,
  \eql{thzf}$$
where $\om_n^2$ and $d_n$ are the eigenvalues and degeneracies of minus the
Laplacian, $-\De_2$, on the spatial section, $\man$. The dash means that
the denominator should never be zero.

The free energy (effective Lagrangian) is obtained from the limit
  $$
  F={i\over2}\lim_{s\to0}{\ze(s,\be)\over s}\,,
  \eql{eff}$$
which, in general, can have infinities. Leaving these aside, in
[\pref{Dow4}] the question of inversion symmetry was discussed directly,
and most conveniently, from (\peq{thzf}) as displaying most clearly any
symmetry under interchange of $\be$ and a scale $a$, which is embedded in
the $\om_n$, typically $\om_n\sim 1/a$.

It is possible to write the formalism in several equivalent forms often
characterised by the way in which the mode information is organised. Each
form has its own features. In [\pref{Dow4}] two such were outlined and it
is the one which starts from the familiar statistical sum that we wish to
concentrate on here.

In order to avoid certain problems, we deal only with internal energies,
and leave for future work a treatment of the free energy and a complete
thermodynamical scheme.

As shown in [\pref{DandKe}], the limit (\peq{eff}) leads, or can lead, to
the expression for the total internal energy, $E=\pa(\be F)/\pa\be$,
  $$
  E=E_0+d_0 T+\sum_{\om_n\ne0}{d_n\,\om_n\over e^{\be\om_n}-1}\,.
  \eql{inten}$$
$E_0$ is the zero temperature value and may be the result of a
regularisation/ renormalisation procedure. The effect  of $d_0$ zero modes
has also been included as the second term [\pref{Dow1,Dow2}]. It is this
form we wish to dwell on for a while as it allows one to make contact with
some standard results in elliptic function theory.
\section{\bf 4. Scalars on the odd sphere.}

Information on modes on a sphere can be found in many places going back
many years. For scalars conformal on the space-time, T$\times$S$^d$, the
eigenvalues are perfect squares and the information can be displayed in the
(zero temperature) \zf,
 $$
 \ze(s)={2\over(2r)!}\sum_{n=1}^\infty{(n^2-(r-1)^2)\ldots n^2\over 
 (n^2/a^2)^s}\,,
 \quad d=2r+1,
 \eql{spzet}$$
where, as mentioned in the Introduction, we are focusing on odd spheres.
The degeneracy is a Gegenbauer polynomial. Another way of writing this
involves the Barnes \zf\ but we will not need this here.

Except for the circle, $d=1$, there are no zero modes so for simplicity
let's restrict initially to $d>1$ and make the common expansion of the
degeneracy (\eg [\pref{CandW}]),
  $$
  (n^2-(r-1)^2)\ldots n^2=\sum_{t=2}^{r+1} (-1)^{t+r}\mu_t(r)\,n^{2t-2}
  \eql{scdegen}$$
which allows $\ze(s)$ to be written as a sum of Riemann \zfs, as has been
done many times before. Our object is not to investigate this particular
aspect much further but only to say that from this sum, or otherwise, it is
possible to obtain the finite values of the zero temperature vacuum energy,
$E_0$, as a sum of ordinary Bernoulli numbers, a calculation done a long
time ago, and a number of times since. This is not the most compact form,
but it is suitable for our purposes where, as has been indicated, we are
going to treat the terms in (\peq{scdegen}) separately. For the $(2t-2)$th
power we have
  $$
 a E_0(t)={1\over2}\,\ze_R(1-2t)=-{B_{2t}\over 4t}\,,
  $$
according to general theory [\pref{DandKe,DandB}]. (There is no need to use
the Bernoulli number. One could just leave the Riemann \zf, which is
actually better.)

Let us therefore return to (\peq{inten}) and consider the `partial'
internal energies,
  $$
  aE(t)=-{B_{2t}\over 4t}+\sum_{n=1}^\infty{n^{2t-1}\over e^{2\pi n/\xi}-1}\,,
  \eql{inten2}$$
and, in order to produce an expression that looks familiar, define a new
parameter, $q$, by,
  $$
  q=e^{-\pi/\xi}\,,
  \eql{cue}$$
to rewrite the internal energy (\peq{inten2}) as a $q$--series,
  $$
 \ep_t(\xi)\equiv aE(t)=-{B_{2t}\over 4t}+
  \sum_{n=1}^\infty {n^{2t-1}q^{2n}\over1-q^{2n}}\,.
  \eql{inten3}$$

The summation is an example of a Lambert series and occurs in analytic
number and in elliptic function theory. In the former, $q$ usually is the
square of the $q$ here, (\peq{cue}).

The most rapid way of proceeding is to note the standard result that the
combination (\peq{inten3}) is, up to a factor, an Eisenstein series.
Precisely,
  $$
 \ep_t(\xi)=(-1)^t{(2t-1)!\over2(2\pi)^{2t}}\, G_t(1,i/\xi)\,,
  \eql{inten4}$$
where the Eisenstein series is defined by (we are following Hurwitz,
[\pref{Hurwitz,Hurwitz2}], here),
  $$
  G_t(\om_1,\om_2)=\sumdasht{{m_1,m_2\atop=-\infty}}{\infty}{1\over
  \big(m_1\om_1+m_2\om_2\big)^{2t}}\,.
  \eql{Eis}$$

When $t=1$, the torus case, the sums have to be considered more carefully.
This point will be returned to. A useful discussion is given by Rademacher
[\pref{Rad}].

The behaviour of $\ep_t(\xi)$ under inversion is now apparent from
(\peq{inten4}) and the structure of the Eisenstein series (\peq{Eis}), \ie
  $$
 \ep_t(1/\xi)=(-1)^t {1\over\xi^{2t}}\,\ep_t(\xi)\,.
  \eql{ind}$$
The $G_{k}$ are holomorphic modular forms, invariant under SL(2,\oZ) action
on the periods $\om_1,\om_2$. The simple inversion case $\xi\to1/\xi$
produces (\peq{ind}). The more general action contains other information.

The result (\peq{ind}) when re-expressed in terms of the expression in
(\peq{inten3}) is equivalent to the duality relation,
  $$
  \mu^t\sum_{n=1}^\infty {n^{2t-1}\over e^{2\mu n}-1}-
  (-\mu')^t\sum_{n=1}^\infty {n^{2t-1}\over e^{2\mu' n}-1}=
  \big(\mu^t-(-\mu')^t\big){B_{2t}\over4t}\,,\quad \mu\mu'=\pi^2\,,
  \eql{glaisher1}$$
which is given by Ramanujan, typically without proof. It was derived later
by Rao and Ayyar, [\pref{RandA}] and thereafter by many workers including
Malurkar [\pref{Malurkar}] and Hardy [\pref{Hardy}] who use Mellin
transforms. Berndt [\pref{Berndt}] gives some history which we could modify
by saying that, since (\peq{glaisher1}) follows immediately from
(\peq{inten4}), then it is contained in Glaisher, [\pref{Glaisher}] and, by
the same token, appears even earlier in Hurwitz [\pref{Hurwitz,Hurwitz2}].

The Eisenstein series enter into the Laurent expansion of the  Weierstrass
$\wp$--function with periods $\om_1$ and $\om_2$,
 $$
 \wp(u)={1\over u^2}+\sum_{t=2}^\infty (2t-1)\,G_t\,u^{2t-2}\,,
 \eql{weier}$$
which can be regarded as a generating function for the internal energies,
$\ep_t$.

Because of the restrictions of analyticity and periodicity, $\wp$ satisfies
the Weierstrass differential equation, (originally obtained by Eisenstein),
  $$\eqalign{
  \wp'(u)^2&=4\wp(u)^3-60\,G_2(\om_1,\om_2)\wp(u)-140\,G_3(\om_1,\om_2)\cr
  &\equiv 4\wp(u)^3-g_2(\om_1,\om_2)\wp(u)-g_3(\om_1,\om_2)}
  $$
which leads to a recursion formula giving all the $G_t$ as algebraic
polynomials in just the two {\it invariants}, $g_2$ and $g_3$, with
rational coefficients, which is a very remarkable fact. A few examples are
given in [\pref{Hurwitz2}], footnote on p.33 and a larger list in
[\pref{TandM}] IV p.88.

The calculation has been kindly done for us by Rademacher, [\pref{Rad}],
and gives, expressed in the $\ep_t$,
  $$
 \ep_t(\xi)=12{(t-1)(2t-3)\over(2t+1)(t-3)}\sum_{l=2}^{t-2}\comb{2t-4}{2l-2}
 \ep_l(\xi)\,\ep_{t-l}(\xi)\,,\quad t\ge4\,.
  \eql{recurs}$$
This is an identity in $\xi$ and yields several specific identities
generally of arithmetic interest. For example, in the low temperature
limit, $\xi=0$, \ie $q=0$, (\peq{recurs}) reduces to an identity between
Bernoulli numbers, \eg [\pref{Rad}] p.124.

The recursion is more simply expressed in terms of the expansion
coefficients, $T_t\equiv(2t-1)\,G_t$, in (\peq{weier}), \eg
[\pref{Halphen}] p.92, [\pref{TandM}] Vol I, p.176, [\pref{HandC}] p.155,
[\pref{schwarz}] p.11, as
  $$
  T_t={3\over(t-3)(2t+1)}\,\sum_{l=2}^{t-2}T_l\,T_{t-l}\,.
  \eql{recurs1}$$

Examples of the minimal polynomials resulting from (\peq{recurs}) are
  $$\eqalign{
\ep_4=&120\,\ep_2^2\,,\quad\!\ep_5={2^4\,3^2\,5\,7\over11}\,\ep_2
\,\ep_3\,,\quad
 \ep_6={2^3\,3^2\,5^2\,7\over13}\,(96\,\ep_2^3+\ep_3^2),\cr
 \noalign{\vskip5truept}
  &\ep_7=2^8\,3^3\,5^2\,7\, \ep_2^2\,\ep_3\,,
 \quad\! \ep_8={2^8\,3^4\,5^3\,7^2\over17}\, \ep_2\,(44\,\ep_2^3+\ep_3^2)\,.}
  \eql{minpol}$$

The total energy on the $d$--sphere is, according to equations (\peq{inten}),
(\peq{scdegen}) and (\peq{inten2}), a sum of the $\ep_t$,
  $$
  aE_{{\rm S}^d}\equiv \overline E_d(\xi)={2\over(2r)!}
  \sum_{t=1}^r(-1)^{t+r}\mu_t(r)\,\ep_{t+1}(\xi)\,,\quad d=2r+1\,,
  \eql{sphen}$$
and therefore can also be expressed as a polynomial in $\ep_2$ and $\ep_3$,
or, equivalently in the three-- and five--sphere energies, $\overline E_3$
and $\overline E_5$.

This has numerical import since it means that the energies on all
$d$--spheres are known once those on S$^3$ and S$^5$ are and this is true
for all temperatures, including zero. For example, the energies up to the
nine--sphere are,
  $$\eqalign{
  \overline E_3&=\ep_2\cr
  \noalign{\vskip3truept}
  \overline E_5&={1\over12}(\ep_3-\ep_2)\cr
  \noalign{\vskip3truept}
  \overline E_7&={1\over360}(120\, \ep_2^2-5\,\ep_3+4\,\ep_2)\,\cr
  \noalign{\vskip3truept}
  \overline E_9&={1\over221760}(5040\, \ep_2\,\ep_3-18480\,
  \ep_2^2+539\,\ep_3-396\,\ep_2)\,.}
  \eql{ebars}$$

Apart from this, the behaviour of the energy on S$^d$ under the inversion
$\xi\to1/\xi$ is determined from (\peq{sphen}) and the individual
behaviours, (\peq{ind}). In this way we regain the high temperature form
derived generally in [\pref{DandKe}] in terms of the heat-kernel
coefficients. It is another way of introducing, organising and using the
spectral information.

Quantities equivalent to the $\ep_t$ are considered by Ramanujan
[\pref{Raman}] who computes the expressions analogous to (\peq{minpol}) by
a recursion derived purely algebraically.

\section{\bf5. Spinors on the odd sphere.}

The spinor spectral data on the sphere are well known so  we can be brief
and simply display them via the zero-temperature \zf\ on odd spheres,
  $$
  \ze(s)={2^{(d+1)/2}\,(-1)^{(d-1)/2}\over(d-1)!}\sum_{t=0}^{(d-1)/2}
  \sum_{n=0}^\infty
  {\nu_t(n+1/2)^{2t}\over\big((n+1/2)^2/a^2\big)^s}
  \eql{spinzf}$$
where the degeneracy has again been expanded and some re-arrangement of the
starting points made for convenience.

We will again treat the individual powers in the degeneracy separately,
and, putting  some constants aside for ease, at finite temperature the
fermion partial internal energies are defined by,
  $$
   \eta_t(\xi)= -{1\over2}\ze_t(-1/2)+\sum_{n=0}^\infty {(2n+1)^{2t-1}\over
     e^{\pi(2n+1)/\xi}+1}\,,
  \eql{spar}$$
where the first term is the zero temperature part with,
  $$
  \ze_t(s)=2^{2t-2s-2}\ze_R(2s-2t+2,1/2),
  $$
which can be given in terms of Bernoulli numbers, if desired, as
  $$
  -{1\over2}2^{2t-1}\ze_R(1-2t,1/2)=\big(1-2^{2t-1}\big)\, {B_{2t}\over4t}=
  2^{2t-1}{B_{2t}(1/2)\over4t}\,.
  \eql{ztemp}$$

The point in writing it in this particular way is that we can now
immediately appeal to a formula given by Glaisher, [\pref{Glaisher}] p.64,
which reads, after transcribing the notation and reverting to the
double--sided summations,
  $$\eqalign{
  \sum_{n=0}^\infty {(2n+1)^{2t-1} q^{2n+1}\over1+q^{2n+1}}=
   -\big(1-&2^{2t-1}\big){B_{2t}\over4t}\,\,+\cr
  &{(-1)^{t+1}(2t-1)!\over4\pi^{2t}}\sumdasht{{m_1,m_2\atop=-\infty}}
   {\infty}{{(-1)^{m_1+m_2}\over\big(m_1+im_2/\xi\big)^{2t}}}\,,}
  \eql{glaisher2}$$
where $q=e^{-\pi/\xi}$. A glance at (\peq{spar}), (\peq{ztemp}) and
(\peq{glaisher2}) shows that the spinor partial energy is given by a
sign--modulated, doubly twisted Eisenstein series, $H_t$,
  $$
  \eta_t(\xi)={(-1)^{t+1}(2t-1)!\over4\pi^{2t}}\,H_t(1,i/\xi)\,
  \eql{feren}$$
where,
  $$
  H_t(\om_1,\om_2)=\sumdasht{{m_1,m_2\atop=-\infty}}{\infty}{{(-1)^{m_1+m_2}
\over\big(m_1\om_1+m_2\om_2\big)^{2t}}}\,,
  \eql{tweisen}$$
and obviously, therefore, enjoys the same inversion properties as the
scalar quantity, $\ep_t(\xi)$, \ie
  $$
  \eta_t(1/\xi)=(-1)^t {1\over\xi^{2t}}\,\eta_t(\xi)\,.
  \eql{spind}$$
Re-expressed by (\peq{spar}), this result becomes,
  $$
  \mu^t\sum_{n=0}^\infty {(2n+1)^{2t-1}\over e^{(2n+1)\mu}+1}-
  (-\mu')^t\sum_{n=0}^\infty {(2n+1)^{2t-1}\over e^{(2n+1)\mu'}+1}=-
  \big(\mu^t-(-\mu')^t\big)(1-2^{2t-1}){B_{2t}\over4t}\,,
  \eql{glaisher3}$$
for $\mu\mu'=\pi^2$. According to Berndt, [\pref{Berndt2}], this formula is
due to Rao and Ayyar, [\pref{RandA}], but we see that it is contained in
Glaisher, [\pref{Glaisher}] (but not in Hurwitz [\pref{Hurwitz}]).

Without going into details, we note at this point that relations more
general than (\peq{glaisher1}) and (\peq{glaisher3}) hold when $t$ is
allowed to be negative.

As an already discussed example, [\pref{AandD}], take the three-sphere. The
total internal spinor energy is,
  $$
  \overline E^{\rm f}_3(\xi)={1\over2}\big(\eta_2(\xi)-\eta_1(\xi)\big)\,.
  $$
Under inversion,
  $$
  \overline E^{\rm f}_3(1/\xi)={1\over2}\bigg({1\over\xi^4}
  \,\eta_2(\xi)+{1\over\xi^2}\,
  \eta_1(\xi)\bigg)\,.
  $$
In this case the heat-kernel expansion terminates at the {\it second} term.

The twisted Eisenstein series, $H_t$, (\peq{tweisen}), appear as
coefficients in the Laurent series for the twisted Weierstrass function,
$\widetilde\wp$, (an `elliptic function of the second kind') defined by,
  $$
  \widetilde\wp(u)={1\over u^2}+\sumdasht{{m_1,m_2\atop=-\infty}}
  {\infty}{\bigg({(-1)^{m_1+m_2}\over\big(u-m_1\om_1-m_2\om_2\big)^2}-
{(-1)^{m_1+m_2}\over\big(m_1\om_1+m_2\om_2\big)^2}\bigg)}\,.
  \eql{tweier}$$
This will appear later in Appendices A and B.

However, in order to derive fermion equations analogous to the recursive
(\peq{ebars}), it seems that Jacobi functions are  more suitable and so
we turn to this description.

\section{\bf6. Elliptic functions and alternative descriptions.}
In this, and the following section we expand on the previous analysis and
also discuss inversion relations using different ingredients.

Hurwitz proves (\peq{inten4}) directly, starting from the partial fraction
expansion for the cotangent. The proof in Rademacher [\pref{Rad}] involves
a slight reorganisation and an application of Lipschitz' formula,
  $$
  {(2\pi)^s\over\Ga(s)}\sum_{m=0}^\infty(m+\al)^{s-1}\,e^{-2\pi z(m+\al)}=
  \sum_{n=-\infty}^\infty {e^{2\pi i n\al}\over(z+ni)^s}\,,
  \eql{Lip}$$
to the $m_1$ summation. Equation (\peq{Lip}) is a duality relation and can
be proved using Poisson summation. There are some restrictions. Thus $\Real
z>0$, $0<\al\le1$ and $\Real s>1$. If $0<\al<1$ then $\Real s>0$.

The result, (\peq{inten4}), is also given in Glaisher, [\pref{Glaisher}]
\S72, and Halphen, [\pref{Halphen}], proved in slightly different manners.
Halphen gives some specific examples on p.446. (There is a minor error of 4
in the third line of his eqn.(78).) See also Hurwitz [\pref{Hurwitz2}]
pp.22,33.

Halphen employs Weierstrass' $\wp$, whereas Glaisher develops everything
from the $q$--series for Jacobi elliptic functions. The ones relevant here
for bosons and fermions are those for zs and ds, the former being
Glaisher's notation for the  Jacobi \zf, $Z_1$. The following trigonometric
expansions are standard and are due, effectively, to Jacobi;
  $$
  {2K\over\pi}\,\zs\, u=\cot{\pi u\over2K}+
  4\sum_{n=1}^\infty {q^{2n}\over1-q^{2n}}\,\sin{n\pi u\over K}\,,
  \eql{zs}$$
and
  $$
  {2K\over\pi}\,\ds\, u ={1\over\sin\big(\pi u/2K\big)}-
  4 \sum_{n=0}^\infty {q^{2n+1}\over1+q^{2n+1}}\,
  \sin\big((n+1/2){\pi u\over K}\big)\,,
  \eql{ds}$$
valid for $|\Imag u|<\pi\Real K'/2$.

The notation is the usual elliptic one, with,
  $$
   q=e^{-\mu}=e^{-\pi K'/K}\,,\quad q'=e^{-\mu'}=e^{-\pi K/K'}\,,
  $$
and $K, K'$ given by,
  $$
  K(k)=\int_0^1{dz\over\sqrt{(1-z^2)(1-k^2z^2)}}\,,\quad K'=K(k')\,,
  \eql{trans}$$
where $k^2+k'^2=1$.

In this scheme the key fact is that expansions of (\peq{zs}) and (\peq{ds})
in powers of $u$ will produce expressions for the quantities $\ep_t$ and
$\eta_t$, of (\peq{inten3}) and (\peq{spar}), in terms of $k$ and $K$. This
is done by Glaisher who {\it then} proceeds to derive the relation to
Eisenstein series, (\peq{inten4}). Explicit forms are given later. For now
we note the Laurent expansion of ds,
  $$
  \ds\, u={1\over u}+4\sum_{l=1}^\infty{(-1)^l\over(2l-1)!}
  \bigg({\pi\over2K}\bigg)^{2l}\eta_l\,u^{2l-1}\,,
  \eql{dsexp}$$
exhibiting the partial energies, $\eta_l$, as coefficients.

Temperature inversion is just the interchange, $k\leftrightarrow k'$ so
that $K\leftrightarrow K'$ and the effect can be deduced directly from
(\peq{zs}) and (\peq{ds}) because the functions zs and ds reproduce
themselves, \ie
  $$\eqalign{
  \zs(u,k^2)&=-i\,\zs(-iu,k'^2)-{\pi u\over2KK'}\cr
  \ds(u,k^2)&=-i\,\ds(-iu,k'^2)\,.}
  \eql{period1}$$
We might term this the Jacobian approach to inversion, and the one via
Eisenstein series, the Weierstrassian approach.

Some inversion formulae are given in Whittaker and Watson [\pref{WandW}]
p.535 ex.60 derived from the imaginary transformation of sd. See also
Hancock [\pref{Hancock}], p.308, Ex.4.

The expansions (\peq{zs}) and (\peq{ds}) yield, on setting
$k\to k'$, $u\to-iu$,
  $$
  {2K'\over i\pi}\,\zs(-iu,k'^2)=\coth{\pi u\over2K'}
  -4\sum_{n=1}^\infty {q'^{2n}\over1-q'^{2n}}\,\sinh{n\pi u\over K'}
  \eql{zsh}$$
and
  $$
 {2K'\over i\pi}\,\ds(-iu,k'^2)={1\over\sinh\big(\pi u/2K'\big)}+4
  \sum_{n=0}^\infty{q'^{2n+1}\over1+q'^{2n+1}}\,
  \sinh\big((n+1/2){\pi u\over K'}\big)\,,
  \eql{dsh}$$
so that from (\peq{period1}) and (\peq{zsh}) we obtain the general
inversion formula,
  $$\eqalign{
  K&\bigg[\coth{\pi u\over2K'}
  -4\sum_{n=1}^\infty {q'^{2n}\over1-q'^{2n}}\,
  \sinh{n\pi u\over K'}\bigg]\cr
  &\hspace{*****}=K'\bigg[
  \cot{\pi u\over2K}+4\sum_{n=1}^\infty {q^{2n}\over1-q^{2n}}\,
  \sin{n\pi u\over K}\bigg]+u\,,}
  \eql{zsinv}$$
which reproduces (\peq{glaisher1}) on expansion in $u$, after noting that
$\mu=\pi K'/K$ and $\mu'=\pi K/K'$. We have said before that this is
equivalent to the inversion symmetry (\peq{ind}).

As a special case, an elementary consequence of (\peq{ind}) is that
$\ep_t(1)$ vanishes if $t$ is odd, \cf Glaisher [\pref{Glaisher}],
\S\S86-94. (This follows more directly from a consideration of the
Eisenstein series.) It gives the interesting identity for the Bernoulli
numbers,
  $$
  \sum_{n=1}^\infty {n^{2t-1}\over e^{2\pi n}-1}={ B_{2t}\over 4t}
  \,,\quad t= 3,5,\ldots\,,
  \eql{bernid1}$$
rediscovered many times. Watson's proof, [\pref{watson}], involves an
elliptic function expansion for ds$^2$ similar to (\peq{zs}).

When $t=1$ the identity is modified to,
  $$
  \sum_{n=1}^\infty {n\over e^{2\pi n}-1}= {B_{2}\over4}-{1\over8\pi}=
{1\over24}-{1\over8\pi}\,.
  \eql{bernid3}$$

The `extra' contribution of $1/8\pi$ arises from the inhomogeneous part of
the imaginary transformation of the \zf, $\zs$, (\peq{period1}) and
corresponds, in our earlier terminology, to the zero mode on the circle,
S$^1$. For this special case the vacuum energy is,
  $$
   \overline E_1(\xi)=-{B_2\over2}+{\xi\over2\pi}+2\sum_{n=1}^\infty {n\over
   e^{2\pi n/\xi}-1}\,.
  \eql{toren}$$

The identity (\peq{bernid3}) can be derived via the full duality relation
obtained from (\peq{zsinv}),
  $$
  \mu'\bigg({B_2\over2}-2\sum_{n=1}^\infty {nq'^{2n}\over
   1-q'^{2n}}\bigg)=-
\mu\bigg({B_2\over2}-2\sum_{n=1}^\infty {nq^{2n}\over
   1-q^{2n}}\bigg)+{1\over2\pi}\,,
  \eql{tordual}$$
by setting $\mu=\mu'$. Equations (\peq{zsinv}) and (\peq{tordual}) seem to
have been derived first by Schl\"omilch, [\pref{Schlomilch}]. Curiously he
does not produce (\peq{glaisher1}) although it is implicit in his formula.

Expressed in terms of the internal energy, (\peq{tordual}) becomes,
  $$
   \overline E_1(\xi)-{\xi\over2\pi}=-
   \xi^2\overline E_1(1/\xi)\,,
  \eql{torinv}$$
and we see that the zero modes have disappeared from the left hand side.

In terms of the Eisenstein series, (\peq{inten4}), one encounters the
series $G_1$ which formally diverges. Giving this term meaning, as
described in Rademacher, for example, would produce the same result as
above, but we have managed to bypass this analysis. Indeed, from
(\peq{torinv}), by recognising that the quantity possessing simple
inversion properties is $\overline E_1(\xi)-\xi/4\pi$, one can {\it derive}
the expression for $G_1$, [\pref{Rad}] (63.82),
  $$
  G_1(1,i/\xi)=(2\pi)^2\bigg({B_2\over2}-2\sum_{n=1}^\infty {nq^{2n}\over
   1-q^{2n}}+{\xi\over4\pi}\bigg)\,.
  $$

\section{\bf7. Spinors again.}
In this section we discuss the corresponding equations for fermions and
also give results analogous to the recursive formulae (\peq{minpol}) and
(\peq{ebars}) for bosons. To begin with, we present the fermion equivalents
of the inversions (\peq{zsinv}) and (\peq{bernid1}).

The general inversion formula following from  (\peq{ds}), (\peq{dsh}) and
(\peq{period1}),
  $$\eqalign{
   K&\bigg[{1\over \sinh\big(\pi u/2K'\big)}
   +4\sum_{n=0}^\infty {q'^{2n+1}\over1+q'^{2n+1}}\,
   \sinh\big((n+1/2){\pi u\over K'}\big)\bigg]\cr
   &=K'\bigg[
{1\over\sin\big(\pi u/2K\big)}-4\sum_{n=0}^\infty {q^{2n+1}\over1+q^{2n+1}}\,
  \sin\big((n+1/2){\pi u\over K}\big)\bigg]\,,}
  $$
is equivalent to the Fermi-Dirac relation, (\peq{spind}) or
(\peq{glaisher3}). The identity analogous to (\peq{bernid1}) is,
  $$
  \sum_{n=0}^\infty {(2n+1)^{4t+1}\over e^{(2n+1)\pi}+1}=
  {1\over4}\,\big(2^{4t+1}-1\big){ B_{4t+2}\over 2t+1}\,.
  \eql{bernid2}$$
There is no zero mode special case. The absence is typical for spinor
thermodynamical quantities in general.

In order to extract the $\eta_t$ from (\peq{dsexp}), the explicit Laurent
expansion of ds is required. A number of ways of obtaining this are
described in Appendix B. The one most appropriate at the moment depends on
the differential equation satisfied by ds. The point is that the cubic
recursion thence derived in Appendix B allows any $\eta_t$ to be expressed
as polynomials in $\eta_1$ and $\eta_2$. This means that, in this case, the
energies on the circle and three-sphere are sufficient to determine all
energies. The reason for the contrast with scalars is the absence of an
obstructing zero mode.

The equations corresponding to (\peq{minpol}) are,
  $$\eqalign{
  \eta_3&={240\over7}\,\eta_1\big(\eta_2+8\,\eta_1^2\big)\cr
  \noalign{\vskip3truept}
  \eta_4&={40\over3}\big(7\eta_2^2+240\eta_1^2\eta_2+576\eta_1^4\big)\cr
  \noalign{\vskip3truept}
  \eta_5&={3840\over11}\,\eta_1\big(85\,\eta_2^2+960\,
  \eta_1^2\eta_2+4032\,\eta_1^4\big)\,,}
  \eql{spminpol}$$
and the spinor energies are, for example,
  $$\eqalign{
  \overline E^{\rm f}_1 (\xi ) &= 2 \eta_1 ,\cr
  \overline E^{\rm f}_3 (\xi ) &= \frac 1 2 (\eta_2 -\eta _1) ,\cr
  \overline E^{\rm f}_5 (\xi) &= \frac 1
   {48} (9\eta _1 - 10\eta _2 + \eta_3 ) ,}
  \eql{spebars}$$
allowing $\overline E^{\rm f}_5$ to be expressed in terms of $\overline
E^{\rm f}_1 (\xi )$ and $\overline E^{\rm f}_3 (\xi )$, and this may be
continued.
\section{\bf8. Explicit formulae.}

The value $\xi=1$ is the {\it lemniscate case} and for scalars, \eg
[\pref{Halphen}] p.64,
  $$
  \ep_2(1)={\Ga^8(1/4)\over5120\pi^6}\,,\quad \ep_3(1)=0\,.
  $$
In elliptic function terminology, the lemniscate value corresponds to
$K=K'$ and so to a modulus of $1/\sqrt2$.

This is a specially interesting point, but it is an important fact in the
practical application of elliptic functions that, given the value of
$K/K'$, one can calculate the values of $K$ and $K'$ separately, and also
of $k$. The physical significance of this for us is the following.

Expansion of the right--hand sides of (\peq{zs}) and (\peq{ds}) in powers
of $u$ introduces the quantities $\ep_t$ and $\eta_t$, as has already been
used to derive inversion symmetry. However now we see that the left--hand
sides will yield expressions for $\ep_t$ and $\eta_t$ in terms of $k$ and
$K$. These have been given by Glaisher, for example. Hence, knowing the
temperature, \ie $\xi =K/K'$, one can compute the partial energies from
elliptic function numerics. Furthermore, according to our previous
analysis, see (\peq{ebars}) and (\peq{spebars}), only $\ep_2$, $\ep_3$,
$\eta_1$ and $\eta_2$ need to be evaluated in this way in order that all
spheres are encompassed, in principle.

Making use of Glaisher's expressions, for convenience, we find, for bosons,
  $$\eqalign{
  \ep_2&={1\over15\pi^4}(1-k^2+k^4)\,K^4\cr
  \noalign{\vskip3truept}
  \ep_3&={4\over63\pi^6}(2-k^2)(2k^2-1)(1+k^2)\,K^6\,.}
  \eql{scell}$$

As might be expected, the expansions occur elsewhere. The relation between
the Eisenstein series and the elliptic functions follows from the very
basic connection between Weierstrass and Jacobi, \eg [\pref{Halphen}] p.24,
  $$
  \wp\,u={4K^2\over\sn^2(2Ku)}-{4\over3}(1+k^2)K^2\,,
  \eql{weisn}$$
and power series expansion of the right--hand side again yields the $(k,K)$
polynomials in (\peq{scell}).  The expansion of $\wp\, u$ is given by
(\peq{weier}) as a power series in $g_2$ and $g_3$, the standard
expressions for which are, [\pref{Halphen}], pp.59-60, (with a sign error),
[\pref{TandM}] IV p.92, [\pref{Weber}] p.152, [\pref{Hancock}] pp.201, 298,
[\pref{greenhill}], p.57, [\pref{AandL}] p.384,
  $$\eqalign{
  g_2&={4\over3}\la^2(1-k^2+k^4)
  ={4\over3}\la^2\big(1-(kk')^2\big)\cr
  g_3&={8\over27}\la^3(1+k^2)(1-2k^2)(1-k^2/2)
   ={8\over27}\la^3\big(k'^2-k^2\big)
  \big(1+{1\over2}(kk')^2\big)\,,}
  \eql{gees}$$
where $\la$ is defined by
  $$
   \la=\bigg({2K\over\om_1}\bigg)^2\,.
  $$
Note that our periods are $\om_1$ and $\om_2$ and we are setting $\om_1=1$
and $\om_2=iK'/K$. For useful relational facts see [\pref{TandM}] vol.II,
chap.IV, pp.180,207, and [\pref{Halphen}],pp.44-46.

It is interesting to confirm the inversion symmetry directly from the forms
in (\peq{scell}). Write generally $\ep_t(k,K)=F_t(k^2)K^{2t}(k)$. Inversion
is $k^2\to {k'}^2=1-k^2$. Then (Hancock [\pref{Hancock}] p.308, Ex.4),
  $$
   F_t(1-k^2)=(-1)^t\,F_t(k^2)
  \eql{coin}$$
so that,
  $$
  \ep_t(k',K')=(-1)^t\bigg({K\over K'}\bigg)^{2t}\,\ep_t(k,K)\,,
  $$
which is exactly (\peq{ind}).

Turning now to the fermion case, the expressions we need are most easily
read off from the expansion for \ds given in Glaisher, [\pref{Glaisher}],
  $$\eqalign{
  \eta_1&=-{\la\over24\pi^2}\,(1-2k^2)\cr
  \noalign{\vskip3truept}
  \eta_2&={\la^2\over240\pi^4}\,(7+8k^2-8k^4)\,\cr
  \noalign{\vskip3truept}
  \eta_3&=-{\la^3\over504\pi^6}\,(1-2k^2)(16k^4-16k^2+31)\,,}
    \eql{spell}$$
showing, as a simple check, that $\eta_1(1)=0$, $\eta_3(1)=0$ at the
lemniscate point, $k=1/\sqrt2$. The expressions satisfy the relation in
(\peq{spminpol}).

The power series expansions of the elliptic functions used here, although
old and standard, are not emphasised in the usual textbooks, with a few
exceptions. For this reason they are developed a little in Appendix B,
where a recursion for the expansion coefficients of \ds is given and has
been checked against Glaisher's expressions.

\section{\bf 9. Numerical evaluation. Singular moduli.}

Computationally, the nome, $q$, is known from the temperature so the
modulus, $k$, can be calculated, usually via a power series in $q$, and
then $K(k)$ follows straightforwardly by the Legendre-Landen-Gauss method
(\eg Appell and Lacour, [\pref{AandL}] \S198, Hurwitz and Courant,
[\pref{HandC}] p.242, Greenhill [\pref{greenhill}], p.322) or one could,
pragmatically, just use the built-in algorithms of Mathematica.

Generally, there seems no advantage in using an elliptic formulation, as
opposed to a simple evaluation of the defining thermal state summations,
(\peq{inten3}) and (\peq{spar}), taken in conjunction with inversion,
(\peq{ind}) and (\peq{spind}). However, as well as the lemniscate case,
there are many other special values of $K'/K$ that have `exact' expressions
for $k$ so that only the computation of $K$ from $k$ remains and in many
cases, like the lemniscate one, $K$ is known in closed form.

Abel has shown, for $K'/K$ of the special form,
  $$
  {K'\over K}={a+b\sqrt n\over c+d\sqrt n}\,,\quad a,b,c,d,n\in \oZ\,,
  \eql{Abel}$$
that $k$ is the solution of an algebraic equation with integer
coefficients. Relatedly, it is known that this also applies when
$K'/K=\sqrt N$, where $N$ is rational. In particular, when $N$ is an
integer, the modulus, $k\equiv k_N$, is referred to as a {\it singular
modulus}.

There are many classic works on modular equations and singular moduli. A
little known one that gives a direct algebraic construction for prime order
is Russell, [\pref{Russell}]. Lists of $k_N$  can be found, \eg in Weber
[\pref{Weber,Weber2}] and in Greenhill [\pref{greenhill2}]. The notebooks
of Ramanujan also contain many examples of modular equations,
[\pref{BandCh}].

It is also known that the complete elliptic integral, $K$, evaluated at a
singular modulus can be expressed in terms of Gamma functions with rational
arguments. The $N=1,3,4$ cases are classic, (\eg [\pref{WandW}]) and the
$N=2$ case is treated by Glasser and Wood [\pref{GandW}], although it seems
to be implicit in work of Ramanujan.

The general case was discussed by Selberg and Chowla, [\pref{SandC}], who
proved that the complete elliptic integral, $K(k_N)$, is always expressible
in terms of Gamma functions and gave the $N=5$ and $N=7$ forms.

The lowest value, $N=1$, is just the lemniscate value touched upon earlier.
When $N$ is a square, $K(k_N)$ is an algebraic number times $K(k_1)$. More
generally when $N$ contains a square, $K(k_N)$ is an algebraic number times
$K(k_{N_0})$ where $N_0$ is the square--free part of $N$.

Much effort has gone into finding $K(k_N)$ together with $k_N$ and, by now,
many specific examples have been listed \eg
[\pref{GandW,SandC,zucker,zucker2,ZandR,JandZ1,JandZ2,BandZ,BandB}].
Borwein and Zucker, [\pref{BandZ}], show that using the beta function often
provides simplified expressions.

In the following we calculate closed forms of the energy in one, three and
five dimensions for a selection of values, $\xi_N$, of $\xi$ that
correspond to singular moduli.

Examples we will look at are $N=1,2,3,6,7,10,15$, all of which can be found
in Zucker and Joyce, [\pref{JandZ2}], for example. For convenience we
summarize the singular moduli used,
  $$ \eqalign{ k_1 &= \frac 1 {\sqrt 2}\cr
 k_2 &= \sqrt 2 -1 \cr
k_3 &= \frac 1 4 \sqrt 2 ( \sqrt 3 -1)\cr
k_6 &=(\sqrt 3 -\sqrt 2 ) (2-\sqrt 3)\cr
k_7 &= \frac 1 8
\sqrt 2 (3-\sqrt 7)\cr
 k_{10} &= (\sqrt{10} -3) (\sqrt 2 -1)^2\cr
k_{15}&= \frac 1 {8\sqrt 2} (2-\sqrt 3)(\sqrt 5 -\sqrt 3)(3-\sqrt 5),}  
  $$
and for $K(k_N)$, correspondingly,
  $$\eqalign{
K(k_1) &= \frac 1 4 \beta \left(1/4\right) \cr
K(k_2) &= \frac{2^{3/4}}{16} \beta \left(1 /8 \right)\cr
K(k_3)&= \frac{2^{1/3} 3^{1/4}}{12} \beta \left(1/ 6 \right)\cr
K(k_6)&= \frac{2^{1/12}3^{1/4}}{48} (\sqrt 2 -1) (\sqrt 3 +1) \beta
\left(1/24 \right) \cr
K(k_7) &= \frac{2^{5/7}7^{3/4}}{14} \frac{\beta \left( 1 /7 \right)
 \beta \left(2 /7 \right)}{\beta \left(1/14 \right)}\cr
K(k_{10}) &=\frac{10^{1/4}}{80} (\sqrt5-2)^{1/2}
\frac{\beta\left(1/40\right)\beta\left(9/40
\right)}{\beta\left(3/8 \right) } \cr
K(k_{15})&= \frac{15^{1/4}}{60} (\sqrt 5 -1 )\, \frac{\beta \left( 1/
15\right) \beta\left(4/15\right)}{\beta\left(1/3\right)}\,, }
  $$
where the beta function,
  $$ \beta (x) \equiv B(x,x) =
\frac{\Gamma^2 (x) }{\Gamma (2x)}\,,
  $$
has been used.

The energies can be calculated at the values $\xi=\xi_N=1/\sqrt N$. We give
below the first three values, the simpler remaining ones being relegated to
Appendix C. The scalar closed expressions in three dimensions are, from
(\peq{ebars}) and (\peq{scell}),
  $$
\eqalign{
\overline E_3 (\xi_1)&= \frac{{\Gamma^8({1}/{4})}}{5120\,{\pi}^6}
\cr
\noalign{\vskip3truept}
\overline E_3 (\xi_2)&=
 {\left( 3 - 2\,\sqrt2\right)\over3072\,\pi^2}\, {{\Gamma^4({1}/{8})}
 \over {\Gamma^4({5}/{8})}}\cr
 \noalign{\vskip3truept}
\overline E_3 (\xi_3)&= \frac{{\Gamma^4({1}/{6})}\,
{\Gamma^4({1}/{3})}}{12288\,{\pi^6 }}\,.}
 $$
and in five dimensions,
  $$\eqalign{
  \overline E_5 (\xi_1)&=-
\frac{{\Gamma^8(\frac{1}{4})}}{61440\,{\pi }^6}\cr
\noalign{\vskip3truept}
\overline E_5 (\xi_2)&=
\frac{{\Gamma^8(\frac{1}{8})}\,
     \left( \left( 7 - 5\,{\sqrt{2}} \right) \,
        {\Gamma^4(\frac{1}{8})} +
       24\,\left( -3 + 2\,{\sqrt{2}} \right) \,{\pi }^2\,
        {\Gamma^2(\frac{1}{4})} \right) }{7077888\,{\pi }^6\,
     {\Gamma^6(\frac{1}{4})}}\cr
     \noalign{\vskip3truept}
\overline E_5 (\xi_3)&=-
 \frac{\left( {\Gamma^6(\frac{1}{6})}\,
       \left( 56\,2^{\frac{2}{3}}\,{\pi }^4 +
         11\,{\Gamma^6(\frac{1}{3})} \right)  \right) }{24772608\,
     {\pi }^9}\,.}
  $$

The spinor energies are given by (\peq{spebars}).
In one dimension,
  $$\eqalign{
   \overline E^{\rm f} _1 (\xi_1 ) &= 0      \cr
   \noalign{\vskip5truept}
   \overline E^{\rm f} _1 (\xi_2 ) &=
\frac{\left( 5 - 4\,{\sqrt{2}} \right)
         \,{{\Gamma}^2(\frac{1}{8})}}
   {96\,\pi \,{{\Gamma}^2(\frac{5}{8})}}
      \cr
      \noalign{\vskip5truept}
\overline E^{\rm f} _1 (\xi_3 ) &=
   -\frac{\left( {{\Gamma}^2(\frac{1}{6})}\,
       {{\Gamma}^2(\frac{1}{3})} \right) }{96\,{\pi }^3}\,,}
       $$
and in three,
  $$\eqalign{
\overline E^{\rm f}_3 (\xi_{1}) &=
\frac{3\,{\Gamma^8(\frac{1}{4})}}{2560\,{\pi }^6}\cr
\noalign{\vskip5truept}
\overline E^{\rm f}_3(\xi_{2}) &= \frac{{\Gamma^4(\frac{1}{8})}\,
     \left( \left( -21 + 16\,{\sqrt{2}} \right) \,
        {\Gamma^4(\frac{1}{8})} +
       32\,\left( 8 - 5\,{\sqrt{2}} \right) \,{\pi }^2\,
        {\Gamma^2(\frac{1}{4})} \right) }{49152\,{\pi }^4\,
     {\Gamma^4(\frac{1}{4})}}\cr
     \noalign{\vskip5truept}
\overline E^{\rm f}_3 (\xi_{3}) &=
 {1\over3072\,\pi^6}\bigg({\Gamma^4(\frac{1}{6})}\,
      {\Gamma^4(\frac{1}{3})} +
     \frac{384\,{\pi }^5\,{\Gamma^2(\frac{7}{6})}}
      {{\Gamma^2(\frac{2}{3})}}\bigg)\,.}
  $$

In five dimensions, simpler spinor examples are,
  $$\eqalign{
\overline E^{\rm f}_5 (\xi_{1})
&= -\frac{32\,{ {\Gamma}^8(\frac{5}{4})}}{{\pi }^6}\cr
\overline E^{\rm f}_5 (\xi_{3}) &=
 -\frac{\left( 336\,2^{\frac{1}{3}}\,{\pi }^8 +
       70\,2^{\frac{2}{3}}\,{\pi}^4\,{ {\Gamma}^6(\frac{1}{3})} +
       5\,{ {\Gamma}^{12}(\frac{1}{3})} \right) }{96768\,{\pi}^6\,
     { {\Gamma}^6(\frac{2}{3})}}\,.}
  $$

We now present some numbers computed from the above formulae which,
incidentally, agree well with a direct evaluation using the elliptic
numerical functions available in Mathematica. As a check, for increasing
$N$, \ie decreasing temperature, the values approach the standard Casimir
values, which are usually expressed in Bernoulli numbers. This could be
thought of only as a necessary confirmation that the second terms in
(\peq{inten3}) and (\peq{spar}) must tend to zero as $\xi$ becomes smaller.
However, it does mean that however complicated the closed forms become for
large $N$, they must ultimately tend to a Bernoulli expression \ie a simple
rational number! The zero temperature quantities are completely entangled
in the finite form expression.
\newpage
\noindent For scalars,
\vskip3truept
  $$ \matrix{ \overline E_3 (\xi_1)&=
0.006065678717786288844 & \overline E_5 (\xi_1)
&= -0.0005054732264821907370 \cr
\overline E_3 (\xi_2)&= 0.004305183178876425232 
& \overline E_5 (\xi_2)&=
-0.0005125278116483441533   \cr \overline E_3 (\xi_3)
&= 0.004185448373868254025
& \overline E_5 (\xi_3)&= -0.0005125654321803054872  \cr 
\overline E_3 (\xi_6)&=
0.004166873656512456887 & \overline E_5 (\xi_6)
&= -0.0005125661374804481332 \cr
\overline E_3 (\xi_7)&= 0.004166726978035004716  
& \overline E_5 (\xi_7)&=
-0.0005125661375588626478 \cr \overline E_3 (\xi_{10})
&= 0.004166669015900687631
& \overline E_5 (\xi_{10})&= -0.0005125661375661265283 
\cr \overline E_3 
(\xi_{15})&=
0.004166666693680499456  & \overline E_5 (\xi_{15})
&= -0.0005125661375661375647
\cr \overline E_3(\xi_\infty)        
&= 0.0041\dot6\hspace{***********}&
\overline E_5(\xi_\infty) &
= -0.000512566\dot13756\dot6\hspace{*****}\cr
                   &=1/240\hspace{************} &
                   &=-31/6480.\hspace{***********} }
 $$

\noindent For spinors,
\vskip3truept

   $$\matrix{
\overline E^{\rm f}_3 (\xi_{1}) &= 0.03639407230671773306  &
\overline E^{\rm f}_1 (\xi_{1}) &=  0\hspace{****************}\cr
\overline E^{\rm f}_3 (\xi_{2}) &= 0.03543620656643122772 &
\overline E^{\rm f}_1 (\xi_{2}) &=  -0.06007307862716140024\cr
\overline E^{\rm f}_3 (\xi_{3}) &= 0.03541764326165557577 &
\overline E^{\rm f}_1 (\xi_{3}) &=  -0.07470339906473023869\cr
\overline E^{\rm f}_3 (\xi_{6}) &= 0.03541666779673327755  &
\overline E^{\rm f}_1 (\xi_{6}) &=  -0.08242382467179113065\cr
\overline E^{\rm f}_3 (\xi_{7}) &= 0.03541666684440446497  &
\overline E^{\rm f}_1 (\xi_{7}) &=  -0.08242382467179113065\cr
\overline E^{\rm f}_3 (\xi_{10}) &=  0.03541666666803304403&
\overline E^{\rm f}_1 (\xi_{10}) &= -0.08242382467179113065\cr
\overline E^{\rm f}_3 (\xi_{15}) &= 0.03541666666666835152&
\overline E^{\rm f}_1 (\xi_{15}) &= -0.08332293842073511712\cr
\overline E^{\rm f}_3(\xi_\infty)         &= 0.03541\dot6\hspace
{**********}&
\overline E^{\rm f}_1(\xi_\infty)         &=-0.08\dot3\hspace
{************}\cr
                       &=17/48\hspace{************}&
                       &=-1/12.\hspace{************} }
  $$

These numbers also show that the range, $0\le\xi\le 1$, is essentially a
Casimir range. Inversion can be used to obtain values in the complementary
`high temperature' region, $1\le\xi\le\infty$.

\section{\bf 10. Conclusion and extensions.}
We have presented expressions for the internal energies of conformally
coupled scalars and spinors on odd spheres in terms of elliptic function
quantities. They can be used for convenient numerical evaluation, but our
main aim, which has become somewhat overshadowed, was really to investigate
the question of temperature inversion symmetry. Regarding this, it was only
for the three-sphere that an overall symmetric quantity (the total energy)
could be found. For the others only the `partial energies' satisfied the
symmetry. The same structure is noticed by Kutasov and Larsen,
[\pref{KandL}]. Thus the statements in Cardy, [\pref{Cardy}], while
correct, are, possibly, slightly misleading. See also Oshima
[\pref{Oshima}].

The closed form expressions at temperatures corresponding to singular
moduli are perhaps somewhat academic. They correspond to denominators in
the thermal \zf, [\peq{thzf}], of the form $n^2+N\,m^2$ indicating number
theory involvement, [\pref{Cox}], and hence elliptic functions. Everything
could be obtained from the Epstein form of the thermal \zf. (See related
calculations by Kennedy [\pref{Kennedy}] on S$^3/\Ga$.) Indeed it was by
the use of the known `exact' forms of certain Epstein functions that Zucker
derived values of $K(k_N)$. Most other calculations also use Kronecker's
limit formula, following Selberg and Chowla, [\pref{SandC}]. It thus seems
that our discussion is a little circular but it is systematic and has the
advantage of uniformity of treatment. One must also not forget that there
seem to be purely algebraic ways of obtaining $K(k_N)$, [\pref{BandCh}],
independent of the limit formula. Not only that but it would be otherwise
difficult to treat $K'/K$ of the more general Abel structure (\peq{Abel}).

As further justification of our elliptic approach, we draw attention to the
fact that the internal energy on the arbitrary odd sphere, S$^d$, is a
polynomial in just two quantities, see (\peq{ebars}) and (\peq{spebars}).

For scalars this is a simple consequence of standard analytical properties
of the Weierstrass function, $\wp$, but seems not so obvious in the Epstein
formulation.

For spinors, the explicit construction of a cubic recursion seems to be
new. It is most easily derived directly from the differential equation
satisfied by ds.

There are a large number of relations that we have not brought into the
present discussion. For example, the double sum representations of
thermodynamic quantities obtained in the earlier work [\pref{DandC}] and
[\pref{AandD}] can be manipulated. One of the summations can be performed
yielding a sum of inverse powers of hyperbolic functions. These constitute
adequate numerical expressions and also can be turned into elliptic
function form to make contact with the present approach, \cf
[\pref{KandS}].

Regarding even spheres, no direct relation to elliptic functions could be
detected. The sums relevant for standard thermodynamic scalars (bosons) are
  $$
  \sum_{n=0}^\infty (2n+1)^{2t}{q^{2n+1}\over1-q^{2n+1}}.
  $$
An obvious generating function, in the spirit of odd spheres, is
  $$
   \sum_{n=0}^\infty {q^{2n+1}\over1-q^{2n+1}}\cos\big((n+1/2)
  {\pi u\over K}\big)
  $$
with periodicity $4K$, but no periodicity in the $iK'$ direction. As it
stands, this makes a connection with Jacobian elliptic functions impossible
and again indicates the necessity of altering the thermal periodicity. The
same statements hold for spinors. In future work we intend to investigate
this question.

We will also investigate a general elliptic description of the complete
thermodynamics of the system, not just the internal energy.

Finally, regarding references, we have not been able to consult some
classic texts, such as those  by Enneper, Dur\`ege and Jordan on elliptic
functions. 
\vskip 20truept
{\bf Acknowledgment:} KK was supported by the Max-Planck-Institute for
Mathematics in the Sciences, Leipzig.
 
\newpage

\section{\bf Appendix A. Fermion expressions.}

As might be expected in a topic involving elliptic functions there are many
routes to required results. This will be illustrated in this appendix by
relating fermion and boson quantities. While not absolutely necessary, the
analysis usefully invokes several basic pieces of elliptic technology.
There are various ways of writing and re-arranging the fermion quantities.

In terms of the Weierstrass approach via the twisted $\widetilde\wp$,
(\peq{tweier}), one can seek to bring back the ordinary Weierstrass
function, $\wp$, by re-arranging the summation lattice in (\peq{tweier}).
If this is drawn out, it is seen that the summations divide into a positive
and a negative square lattice, displaced, one from the other, by, say,
$\om_1$ along the real axis and both rotated by 45$^\circ$ with respect to
the original $m_1,m_2$ lattice. Therefore define new summation integers by
$M=m_1+m_2$ and $m=m_1-m_2$. The positive lattice corresponds to $M$ and
$m$ both being even, and the negative lattice to $M$ and $m$ both odd.
Simple algebra then gives,
  $$
  \widetilde\wp(u;\om_1,\om_2)=\wp(u;\widetilde\om_1,\widetilde\om_2)-
  \wp(u+\om_1;\widetilde\om_1,\widetilde\om_2)
+ \wp(\om_1;\widetilde\om_1,\widetilde\om_2)\,,
  \eql{fertrans}$$
where
  $$\eqalign{
  \widetilde\om_1&=\om_1-\om_2\cr
  \widetilde\om_2&=\om_1+\om_2}
  \eql{srot}$$
so that $\om_1=(\widetilde\om_2+\widetilde\om_1)/2$ is a half-period of
$\wp(u;\widetilde \om_1,\widetilde\om_2)$. This function has a negative
discriminant. The scaled rotation, (\peq{srot}), has determinant 2.

The same transformation can be applied to the normal $\wp$--function, to
give
  $$
  \wp(u;\om_1,\om_2)=\wp(u;\widetilde\om_1,\widetilde\om_2)+
  \wp(u+\om_1;\widetilde\om_1,\widetilde\om_2)
- \wp(\om_1;\widetilde\om_1,\widetilde\om_2)\,.
  \eql{wptrans}$$
Combined with (\peq{fertrans}) this yields,
  $$
  \widetilde\wp(u;\om_1,\om_2)=2\wp(u;\widetilde\om_1,\widetilde\om_2)-
  \wp(u;\om_1,\om_2)\,,
  \eql{fertrans2}$$
which can be used to give an expansion of the fermion $\widetilde\wp \,u$
in terms of that for $\wp\,u$. Thus, similarly to (\peq{weier}), we write,
  $$
 \wp(u;\widetilde\om_1,\widetilde\om_2)
 ={1\over u^2}+\sum_{t=2}^\infty (2t-1)\,\widetilde G_t\,u^{2t-2}\,,
 \eql{feweier}$$
so that,
  $$
 \widetilde\wp(u;\om_1,\om_2)={1\over u^2}+\sum_{t=2}^\infty (2t-1)\,
 \big(2\widetilde G_t-G_t\big)\,u^{2t-2}\,,
 \eql{feweier2}$$
and therefore
  $$
  H_t=2\widetilde G_t-G_t\,,
  \eql{ht}$$
allowing $\eta_t$ to be found from (\peq{feren}).
This will be taken further in Appendix B.

Relation (\peq{fertrans2}) expresses the fermion function, $\widetilde\wp$,
in terms of the boson one, $\wp$, which is sometimes useful.

Introducing yet another, more flexible notation, equivalent to the \zs
function is the theta function construction (\cf [\pref{TandM}], vol.III
p.117),
  $$
  2K\zs(2Ku)\equiv\Phi(u,\tau)={d\over du} \log\th_1(u,\tau)\,,
  \eql{theta1}$$
with
  $$
  q=e^{i\pi\tau}=e^{-\mu}=e^{-\pi K'/K}.
  $$
Thomae denotes $\Phi$ by $Z_{11}$, [\pref{Hancock}] p.295.

In terms of $\Phi$, the fermion function, ds, can be written formally as
the difference of two composite  expressions, (\cf [\pref{Rad}], \S\S84,
85),
  $$\eqalign{
  2K\,\ds(2Ku)&= {1\over2}\Phi\big(u/2,(1+\tau)/2\big)-{1\over2}
  \Phi\big((u+1)/2,(1+\tau)/2\big)\cr
  \noalign{\vskip3truept}
  &={d\over du}
  \log\,{\th_1(u/2,(1+\tau)/2)\over
  \th_2(u/2,(1+\tau)/2)}\cr
  \noalign{\vskip3truept}
  &={d\over du}\log\,{\th_1(u/2,\tau)\,\th_3(u/2,\tau)\over
  \th_2(u/2,\tau)\,\th_0(u/2,\tau)}\,.}
  \eql{theta2}$$
The last equality is a consequence of the quadratic Gauss transformation
expressed in theta functions, (\eg [\pref{TandM}] II, pp.119,120,207). This
can be confirmed by combining the relevant $q$--series by eye (\eg
[\pref{TandM}] III,p.117, [\pref{Halphen}], p.431, [\pref{Glaisher}]) and
this leads onto another way of relating fermion and boson expressions which
is to start from the left-hand side of the relation (\peq{glaisher2}), and
work at the $q$-series level by applying, {\it \`a la} Jacobi, the trivial,
and common, identity,
  $$
  {x\over1+x}={x\over1-x}-{2x^2\over1-x^2}
  \eql{ident1}$$
to the quantities $\eta_t(\xi)$ of (\peq{spar}), which we can write as
  $$
  \eta_t(\xi)=2^{2t-1}{B_{2t}(1/2)\over4t}+\sum_{n=0}^\infty
  {(2n+1)^{2t-1} q^{2n+1}\over1+q^{2n+1}}\,.
  $$
The sum over odds is written as,
  $$
  \sum_{n=0}^\infty {(2n+1)^{2t-1} q^{2n+1}\over1+q^{2n+1}}=
\sum_{n=1}^\infty {n^{2t-1} q^n\over1+q^n}-
\sum_{n=1}^\infty {(2n)^{2t-1} q^{2n}\over1+q^{2n}}\,,
  $$
and (\peq{ident1}) used on both parts,  to give for the right-hand side,
  $$
  \sum_{n=1}^\infty {n^{2t-1} q^n\over1-q^n}-
\sum_{n=1}^\infty {(2n)^{2t-1} q^{2n}\over1-q^{2n}}-
2\sum_{n=1}^\infty {n^{2t-1} q^{2n}\over1-q^{2n}}+
2\sum_{n=1}^\infty {(2n)^{2t-1} q^{4n}\over1-q^{4n}}\,.
  $$
Then, simple algebra, yields the relation,
  $$\eqalign{
  \eta_t(\xi)&=\ep_t(2\xi)-2\ep_t(\xi)-2^{2t-1}\big(\ep_t(\xi)
  -2\ep_t(\xi/2)\big)\cr
  &=\Om_t(2\xi)-2\Om_t(\xi)\cr
  &=\Xi_t(\xi)-2(1+2^{2t-2})\,\ep_t(\xi)\,,}
  \eql{spbos}$$
where $\Om_t(\xi)=\ep_t(\xi)-2^{2t-1}\ep_t(\xi/2)$ and
$\Xi(\xi)=\ep_t(2\xi)+2^{2t}\ep_t(\xi/2)$. The combination (\peq{spbos})
possesses the correct inversion behaviour, as a check. The two parts on the
last line possess the inversion property separately. In terms of $q$, the
right-hand side involves $q^2$, $q^{1/2}$ and $q$.

This expression can also be obtained by reorganising the summations in
the Eisenstein series, $H_{2k}$, and using the basic equation, (\peq{feren}).

In the lowest case of the circle, $t=1$, it is easily confirmed that the
zero mode element of the scalar energy typically cancels on the right-hand
side to leave {\it exactly} the same combination for the total
fermion energy, $\overline
E_1^{\,\,\rm f}$,
  $$
  \overline E_1^{\,\,\rm f}(\xi)=\overline E_1(2\xi)-4\overline E_1(\xi)
  +4\overline E_1(\xi/2)\,.
  \eql{fercirc}$$

Some relevant calculations on the relation between $\widetilde\wp$ and $\wp$
can be found in Tannery and Molk,
[\pref{TandM}] IV, pp.37-40. In our notation they have,
  $$
  \wp(u;\om_1,\om_2)=\wp(u;\widetilde\om_1,\widetilde\om_2)-
{(\widetilde e_2-\widetilde e_1)(\widetilde e_2-\widetilde e_3)\over
\wp(u;\widetilde\om_1,\widetilde\om_2)-\widetilde e_2}\,,
  \eql{tandm40}$$
which also follows from (\peq{fertrans}), and leads to another form for
$\widetilde\wp$,
  $$
  \widetilde\wp(u;\om_1,\om_2)=
\wp(u;\widetilde\om_1,\widetilde\om_2)+
{(\widetilde e_2-\widetilde e_1)(\widetilde e_2-\widetilde e_3)\over
\wp(u;\widetilde\om_1,\widetilde\om_2)-\widetilde e_2}\,.
  \eql{spwp}$$

We have not used this although it does yield the required
expansions after, say, the introduction of two Jacobi functions, \eg
$\widetilde{\sn}^2$ and $\widetilde\dc^2$.

It is possible to avoid the new modulus, $\widetilde k$, by repeating the
transformation to the periods, $\widetilde \om_1,\widetilde\om_2$. This
effectively amounts to doubling the periods to $2\om_2,2\om_1$, which are
periods of $\widetilde\wp$. The necessary analysis is given by Tannery and
Molk [\pref{TandM}] IV pp.276-280. Their quantities $y,Y$ and $Z$ are, in
our notation,
  $$
  Y=\wp(u;\om_1,\om_2)\,,\quad y=\wp(u;\widetilde\om_1,\widetilde\om_2)\,,
  \quad Z=\wp(u;2\om_1,2\om_2)\,,
  $$
so that the fermion $\widetilde\wp$ is,
  $$
  \widetilde\wp(u;\om_1,\om_2)=2y-Y\,.
  $$
Now $y$ can be expressed rationally in terms of $Z$ which leads to,
  $$
  \widetilde\wp=2Z+2{(\ssse_2-\ssse_1)(\ssse_2-\ssse_3)\over Z-\ssse_2}-Y\,,
  \eql{spbos2}$$
where the $\ssse_i$ are the roots associated with $Z$. By homogeneity
$\ssse_i=e_i/4$ and clearly the modulus associated with $Z$ remains $k$.
Equation (\peq{spbos2}) can provide another way to expand the
$\widetilde\wp$.

As a side remark, since $\widetilde\wp(u)$ has periods $2\om_1,2\om_2$, it
can be expressed rationally in terms of $Z$. We find, using the expression
for $\wp(2u)$ in terms of $\wp(u)$,
 $$
 \widetilde\wp(u;\om_1,\om_2)=-{4Z^4+4\ga_2Z^2+10\ga_3Z+\ga_2^2/4\over
 4Z^3-\ga_2Z-\ga_3}+2{(\ssse_2-\ssse_1)(\ssse_2-\ssse_3)\over Z-\ssse_2}
 $$
where $\ga_2$ and $\ga_3$ are the invariants associated with $Z$. Homogeneity
allows everything to be written in terms of
$\wp_\circ\equiv\wp(u/2;\om_1,\om_2)$. Thus
  $$
  Z(u)={1\over4}\wp_\circ\,,
  \quad \ga_2={1\over16}g_2\,,
  \quad \ga_3={1\over64}g_3\,,
  $$
and so
  $$
  \widetilde\wp(u;\om_1,\om_2)=-{1\over4}
  {4\wp_\circ^4+4g_2\wp_\circ^2+10g_3\wp_\circ+{g_2}^2/4\over
 4\wp_\circ^3-g_2\wp_\circ-g_3}+{(e_2-e_1)(e_2-e_3)\over 2(\wp_\circ-e_2)}\,.
  \eql{spwp2}$$
This does not seem to have any computational advantages.

\section{\bf Appendix B. The power series expansion of elliptic functions.}

For our purposes in this paper, we need the expansions of zs and ds.
Glaisher, for example, lists all the expansions but it might be useful to
present an outline of some derivations and other pertinant information.

The expansion of the elliptic functions \sn $u$ \etc as power series in $u$
formed the subject of Weierstrass' first major work dating to 1840. The
derivation is a technical issue that is not treated in much detail in most
textbooks. Exceptions include Fricke, [\pref{Fricke}] and Briot and Bouquet
[\pref{BandB}].

There are many equivalent approaches depending on the representations of
the elliptic functions employed. For example, one could use the theta form
of \zs, given later, or one could employ the relation between \zs and the
Weierstrass \zf,
  $$
  \zs\, u = \ze u-{\eta u\over K}\,,\quad \zs'u=-\wp\,u+{\eta\over K}
  $$

The calculations for sn, cn and dn are described in Fricke, [\pref{Fricke}]
pp.396+, and  K\"onigsberger, [\pref{Konig}], following Weierstrass. There
are two basic approaches. The first is a direct application of Taylor
series. By successive differentiation, Jacobi showed that, for example,
  $$
  {d^{2\nu} \sn\, u\over du^{2\nu}}=\big(a^{(\nu)}_0+
  a^{(\nu)}_1\sn^2\,u+\ldots+a^{(\nu)}_\nu\sn^{2\nu}\,u
\big)\,\sn\, u
  $$
and
  $$
  {d^{2\nu+1} \sn\, u\over du^{2\nu+1}}=\big(a^{(\nu)}_0+
  3a^{(\nu)}_1\sn^2\,u+\ldots+(2\nu+1)a^{(\nu)}_\nu\sn^{2\nu}\,u
\big)\,\cn\, u\, \dn\, u
  $$
where the $a^{(\nu)}$ are easily found by recursion which shows that they
are integer functions of the squared modulus, $k^2$. Then, using the values
$\sn\,0=\,0$, $\cn\,0=1$ and $\dn\,0=1$, the Taylor series of $\sn\, u$
follows directly. A similar technique holds for \cn and \dn and for the
other elliptic function combinations. Some coefficients are given in
[\pref{TandM}] IV pp.92,94.

Obviously one can just use the series for the three basic functions, sn,
cn and dn, to construct those for the others, say for $\ds=\dn/\sn$, by
further recursion or inversion. One could also proceed directly, as for
sn, using $\ds'^2=(k^2+\ds^2)(\ds^2-k'^2)$.

A related technique, and the one referred to in the main body of this
paper, is to use the derivative relations (we give these for the case we
are interested in)
  $$
  {d\over du}\,\ds\,u=-\cs\,u\cdot\ns\,u\,,\quad
  {d\over du}\,\cs\,u=-\ns\,u\cdot\ds\,u\,,\quad
  {d\over du}\,\ns\,u=-\ds\,u\cdot\cs\,u\,,
  $$
mutually connecting the trio of functions, ds, ns, cs.

The conventional way of extracting the information in these relations
consists of setting up three coupled recursion relations for the expansion
coefficients of the three functions, \eg Fricke [\pref{Fricke}], p.400.
Alternatively, further differentiation yields the differential equation for
ds,
  $$
  ds''u=2ds^3u-(1-2k^2)\,\ds\,u\,,
  $$
and substitution of the expansion (\peq{dsexp}) yields, after a little
algebra, the cubic recursion for $j\ge3$,
  $$\eqalign{
  \eta_j={4(2j-1)!\over(j-2)(2j+1)}\sum_{l=1}^{j-2}&\bigg[
  {3\eta_l\eta_{j-l}\over(2l-1)!(2j-2l-1)!}+\cr
  &\sum_{n=1}^{j-l-1}{4\eta_l\eta_n\eta_{j-l-n}\over
  (2l-1)!(2n-1)!(2j-2l-2n-1)!}\bigg]\,.}
  \eql{cubinv}$$
This is the fermion analogue of the boson recursion, (\peq{recurs}), coming
from $\wp$.

In content, (\peq{cubinv}) is the same as the three  coupled recursions,
but is more compact and clearly shows that all coefficients are determined
by just the first two, $\eta_1$ and $\eta_2$. It is quickly checked that
the results agree with those listed by Glaisher and leads to
(\peq{spminpol}) and (\peq{spell}). It may not be the most efficient way of
determining the highest polynomials, but it is formally instructive.

For completeness we mention the second approach, adopted by Weierstrass,
[\pref{Weierstrass, Weierstrass2}], who writes, following Abel, the
elliptic functions as ratios of power series by representing them as
ratios, effectively of theta functions but more handily of the earlier Abel
functions, $Al_i(u)$, which are just factors different from theta
functions.

One has
  $$
  \sn\, u={Al_1(u)\over Al_0(u)}\,,\quad
  \cn\, u={Al_2(u)\over Al_0(u)}\,,\quad
  \dn\, u={Al_3(u)\over Al_0(u)}\,.
  $$
Weierstrass' next step is to Taylor expand the $Al_i(u)$ in a double power
series in $k^2$ and $u$ using the differential equations that they satisfy.
The coefficients are then determined by recursion. The calculation is
straightforward and well adapted to machine evaluation. The result is the
series,
  $$\eqalign{
  Al_0(u)&=\sum_{m=0}^\infty (-1)^{m-1} C_m {u^{2m}\over(2m)!}
\,,\quad Al_1(u)= \sum_{m=0}^\infty (-1)^m A_m {u^{2m+1}\over(2m+1)!}\cr
  Al_2(u)&=\sum_{m=0}^\infty (-1)^m B_m {u^{2m}\over(2m)!}\,,
\quad Al_3(u)=\sum_{m=0}^\infty (-1)^m D_m {u^{2m}\over(2m)!}\,, }
  $$
with $A_0=B_0=D_0=1$ and $C_0=-1,\,C_1=0$, from which the expansions for
any elliptic function can be found, after a further recursion in the form
of a forward substitution. For example, writing, \cf (\peq{dsexp}),
  $$
  \ds\, u={Al_3(u)\over  Al_1(u)}
  ={1\over u}+\sum_{n=1}^\infty (-1)^n\,d_n\,{u^{2n-1}\over(2n)!}\,,
  \eql{dsexp2}$$
one finds the relation between coefficients,\mgn{CHECK!!}
  $$
    \sum_{i=0}^{r-1}\comb{2r+1}{2i+1}\, d_{r-i}\, A_i=(2r+1)D_r-A_r\,,
  \eql{coerel1}$$
and so
  $$
  (2r+1)\,d_r=(2r+1)D_r-A_r-\sum_{i=1}^{r-1}\comb{2r+1}{2i+1}\,d_{r-i}\,A_i
  \eql{fsub}$$
which is suitable for successive substitution. It is very easy to combine
the recursions by machine, however Weierstrass [\pref{Weierstrass2}] gives
expressions for the $A_m,B_m,C_m,D_m$ coefficients up to $m=10$ and it is a
simple matter to check that (\peq{fsub}) reproduces the results from
(\peq{cubinv}).

  This is an adequate scheme for practical purposes but somewhat utilitarian.
An aesthetically more pleasing way of deriving the necessary series
expansions is to use the equation for the twisted $\wp$--function,
(\peq{feweier2}). In order to write this in terms of the modulus $k$, which
corresponds to the original periods $\om_1,\om_2$, the modulus, $\widetilde
k$, associated with the rotated periods, $\widetilde\om_1,\widetilde\om_2$,
has to be related to $k$, with a corresponding statement holding for the
complete functions, $K$ and $\widetilde K$. The transformation from
$\om_1,\om_2$ to $\widetilde\om_1,\widetilde\om_2$ is a combination of
modular transformations with a Gauss quadratic one, \cf [\pref{TandM}]
II,p.123, but a direct way of finding the relation is the following.

Standard relations in elliptic function theory are
  $$
  k=\sqrt{{e_2-e_3\over e_1-e_3}}\,,\quad
  k'=\sqrt{{e_1-e_2\over e_1-e_3}}\,,\quad{2K\over\om_1}=\sqrt{e_1-e_3}\,,
  $$
where the half-period values are,
  $$
  e_1=\wp\big({\,\om_1\over2};\om_1,\om_2\big)\,\quad
  e_2=\wp\big({\,\om_1+\om_2\over2};\om_1,\om_2\big)\,\quad
  e_3=\wp\big({\,\om_2\over2};\om_1,\om_2\big)\,.
  \eql{edef}$$
and so the problem is reduced to finding the relation between the $e_\al$
and the $\widetilde e_\al$. This can be done from the connection
(\peq{wptrans}) by substituting in various values of $u$ and using the
definitions (\peq{edef}). One has immediately that the last term in
(\peq{wptrans}) is just $\widetilde e_2$. Putting $u$ equal to $\om_1/2$,
$\om_2/2$ and $\om_3/2$ in turn in (\peq{wptrans}), in the first two cases
the $\wp$ functions on the right are evaluated at {\it quarter} periods.
Using standard results, \eg [\pref{Halphen}] p.54, one quickly finds,
  $$\eqalign{
   e_1&=\widetilde e_2-2i\sqrt{\widetilde e_2-\widetilde e_3}
    \sqrt{\widetilde e_1-\widetilde e_2}\cr
  e_3&=\widetilde e_2+2i\sqrt{\widetilde e_2-\widetilde e_3}
    \sqrt{\widetilde e_1-\widetilde e_2}\cr
  e_2&=\widetilde e_1+  \widetilde e_3-\widetilde e_2=-2\,\widetilde e_2\,,}
  \eql{erels}$$
and it is then easy to show that the moduli and complete integrals
are related by
  $$
  2\widetilde k\,\widetilde k'={1\over2kk'}\,,\qquad
  \widetilde\la=
   ikk'\la\,.
  \eql{modrel}$$

It is to be remarked that the new modulus, $\widetilde k$, is complex and
that $\widetilde k'=\widetilde k^*$ so $\widetilde k^2=1/2+ib$ with $b$
real.

It is straightforward to compute the fermion partial energies from
(\peq{feren}) and (\peq{ht}).
According to (\peq{ht }) one needs the combination $2\widetilde g_2-g_2$
which, after using (\peq{modrel}), gives,
  $$
  2\widetilde g_2-g_2=-{\la^2\over6}\big(8\,(kk')^2+7\big)\,.
  \eql{spg2}$$
This is the same as in (\peq{spell}), up to a factor. To deal with $g_3$
we write it in the form given by Ramanujan [\pref{Raman}] eqn.(60),
  $$
  g_3={8\over27}\la^3\big(1+{1\over2}(kk')^2\big)\sqrt{1-(2kk')^2}\,.
  \eql{ramg3}$$
This enables the transformations (\peq{modrel}) to be simply applied
in order to evaluate $2\widetilde g_3-g_3$ which again reproduces the
expression in (\peq{spell}). Note the useful symmetry
  $$
  {i\over2\widetilde k\,\widetilde k'}\sqrt{1-(2\widetilde 
k\widetilde k')^2}=  \sqrt{1-(2kk')^2}\,.
  $$

{}From their structure, the quantities $\ep_t$ and $\eta_t$ will contain a
polynomial in $2kk'$ for $t$ even and a similar polynomial times
$\sqrt{1-(2kk')^2}$ for $t$ odd. This is actually very handy when
evaluating at singular moduli because the combination, $2kk'$, is often the
most convenient one to emerge from the modular equation, \eg [\pref{Russell}].

In terms of the expansion coefficients in (\peq{dsexp2}), $T$ inversion,
(\peq{period1}), implies (compare with the scalar case, (\peq{coin})),
  $$
  d_n(k^2)=(-1)^n\,d_n(1-k^2)\,,
  $$
and so a more appropriate variable would be $1/2-k^2\equiv\ka$. For odd
$n$, $d_n$ vanishes only at $\ka=0$ while if $n$ is even, $d_n$ has a
minimum at $\ka=0$, maxima at $\ka=\pm\ka_0$ and passes through zero at
$\ka=\pm\ka_1$, where $\ka_1\to1.5$ as $n\to\infty$.

Incidentally, purely as a check with Glaisher, using the modular
transformations generated by $T:k\to k'$ and $S:k'\to1/k'$, the expansion
coefficients for the functions \cs and \ns are given in terms of the $d_n$.
Those for cs are $k^{2n}\,d_n(1/k^2)$ and those for ns,
$k'^{2n}\,d_n(-k^2/k'^2)$. This means that the $d_n$ coefficients
could also have been obtained by inverting the series for sn.

Gudermann (Weierstrass' teacher) in 1839 produced expansions which used the
transformation relations for the elliptic functions (see
[{\pref{Hancock}]). Hermite devised a related technique which is outlined
in Briot and Bouquet [\pref{BandB}], see also [\pref{TandM}] II p.209. One
should also mention the work of Andr\'e, [\pref{Andre}]. More recently, a
combinatorial formulation of the coefficients has been employed (\eg Dumont
[\pref{Dumont}]) which is more efficient for the highest coefficients.

\newpage
\section{\bf Appendix C. Internal energies.}
In this appendix we display the remaining simpler singular case
computations of the internal energies in three and five dimenions for
scalars and spinors. This will give an indication of the complexity
encountered, but the results are, at least, in finite terms.

First for scalars, in three dimensions,
  $$
\eqalign{
\overline E_3 (\xi_6)&=
 \frac{\left( 35 - 20\,{\sqrt{2}} + 16\,{\sqrt{3}} - 14\,{\sqrt{6}}
\right)
       \,{\Gamma^4({1}/{24})}}{138240\,{\pi }^2\,
     {\Gamma^4({13}/{24})}}\cr
\overline E_3 (\xi_7)&=
  \frac{17\,{\Gamma^4({1}/{7})}\,
     {\Gamma^4({2}/{7})}\,{\Gamma^4({4}/{7})}}
     {458752\,{\pi }^8}\cr
\overline E_3 (\xi_{10})&=
 \frac{4\,\left( 7725 - 5460\,{\sqrt{2}} + 3452\,{\sqrt{5}} -
       2442\,{\sqrt{10}} \right) \,{\Gamma^4({9}/{40})}\,
     {\Gamma^4({7}/{8})}\,{\Gamma^4({41}/{40})}}
     {{\pi }^2\,{\Gamma^4({3}/{8})}\,
     {\Gamma^4({21}/{40})}\,
     {\Gamma^4({29}/{40})}}\cr
\overline E_3 (\xi_{15})&= \frac{\left( 245 - 104\,{\sqrt{5}} \right) \,
     {\Gamma^4({1}/{30})}\,{\Gamma^8({4}/{15})}}
     {8640000\,2^{{3}/{5}}\,{\pi }^2\,{\Gamma^4({1}/{6})}\,
     {\Gamma^4({1}/{3})}\,{\Gamma^4({17}/{30})}}, }
  $$
and in five,
  $$\eqalign{
\overline E_5 (\xi_7)&=
 -\frac{\left( 3808\,{\pi^4 }\,{\Gamma^4(\frac{1}{7})}\,
        {\Gamma^4(\frac{2}{7})}\,{\Gamma^4(\frac{4}{7})}
        + 171\,{\Gamma^6(\frac{1}{7})}\,
        {\Gamma^6(\frac{2}{7})}\,{\Gamma^6(\frac{4}{7})}
       \right) }{1233125376\,{\pi }^{12}}}
  $$
The corresponding spinor expressions are

   $$\eqalign{
\overline E^{\rm f} _1 (\xi_6 ) &=
    -\frac{2\,\left( 1 + 2\,{\sqrt{2}} - 2\,{\sqrt{3}} \right) \,
     {{\Gamma}^2(\frac{25}{24})}}{\pi \,
     {{\Gamma}^2(\frac{13}{24})}}
     \cr
\overline E^{\rm f} _1 (\xi_7 ) &= -\frac{\left(
{{\Gamma}^2(\frac{1}{7})}\,{{\Gamma}^2
      (\frac{2}{7})}\,{{\Gamma}^2(\frac{4}{7})}
       \right) }{128\,{\pi }^4}
      \cr
\overline E^{\rm f} _1 (\xi_{10} ) &=
     -\frac{\left( 32 - 23\,{\sqrt{2}} +
       4\,{\sqrt{\frac{1273}{10} - 90\,{\sqrt{2}}}} \right) \,
     {{\Gamma}^4(\frac{1}{40})}\,
     {{\Gamma}^4(\frac{9}{40})}
     \,{{\Gamma}^2(\frac{3}{4})}}
     {1280\,{\pi }^2\,{{\Gamma}^2(\frac{1}{20})}\,
     {{\Gamma}^4(\frac{3}{8})}\,
     {{\Gamma}^2(\frac{9}{20})}}
      \cr
\overline E^{\rm f} _1 (\xi_{15} ) &=
    -\frac{\left( 4 - {\sqrt{5}} \right)
    \,{{\Gamma}^2(\frac{1}{30})}\,
{{\Gamma}^4(\frac{4}{15})}}{360\,2^{\frac{4}{5}}\,{\sqrt{5}}\,
     \pi \,{{\Gamma}^2(\frac{1}{6})}\,
     {{\Gamma}^2(\frac{1}{3})}
      \,{{\Gamma}^2(\frac{17}{30})}}
     }
  $$
and
  $$\eqalign{
\overline E^{\rm f}_3 (\xi_{7}) &= \frac{224\,{\pi }^4\,{\Gamma^2
(\frac{1}{7})}\,
      {\Gamma^2(\frac{2}{7})}\,{\Gamma^2(\frac{4}{7})} +
      15\,{\Gamma^4(\frac{1}{7})}\,
      {\Gamma^4(\frac{2}{7})}\,{\Gamma^4(\frac{4}{7})}}
     {114688\,{\pi }^8}\,.}
  $$

\newpage
\section{\bf References.}
\begin{putreferences}
 \ref{BMO}{Brevik,I., Milton,K.A. and Odintsov, S.D. {\it Entropy bounds in
$R\times S^3$ geometries}. hep-th/0202048.}
 \ref{KandL}{Kutasov,D. and Larsen,F. {\it JHEP} 0101 (2001) 1.}
 \ref{KPS}{Klemm,D., Petkou,A.C. and Siopsis {\it Entropy
bounds, monoticity properties and scaling in CFT's}. hep-th/0101076.}
\ref{DandC}{Dowker,J.S. and Critchley,R. \prD{15}{1976}{1484}.}
\ref{AandD}{Al'taie, M.B. and Dowker, J.S. \prD{18}{1978}{3557}.}
\ref{Dow1}{Dowker,J.S. \prD{37}{1988}{558}.}
\ref{Dow3}{Dowker,J.S. \prD{28}{1983}{3013}.}
\ref{DandK}{Dowker,J.S. and Kennedy,G. \jpa{}{1978}{}.}
\ref{Dow2}{Dowker,J.S. \cqg{1}{1984}{359}.}
\ref{DandK}{Dowker,J.S. and Kirsten, K.{\it Comm. in Anal. and Geom.
}{\bf7}(1999) 641.}
\ref{DandKe}{Dowker,J.S. and Kennedy,G. \jpa{11}{1978}{895}.}
\ref{Gibbons}{Gibbons,G.W. \pl{60A}{1977}{385}.}
\ref{Cardy}{Cardy,J.L. \np{366}{1991}{403}.}
\ref{ChandD}{Chang,P. and Dowker,J.S. \np{395}{1993}{407}.}
\ref{DandC2}{Dowker,J.S. and Critchley,R. \prD{13}{1976}{224}.}
\ref{Camporesi}{Camporesi,R. \prp{196}{1990}{1}.}
\ref{BandM}{Brown,L.S. and Maclay,G.J. \pr{184}{1969}{1272}.}
\ref{CandD}{Candelas,P. and Dowker,J.S. \prD{19}{1979}{2902}.}
\ref{Unwin1}{Unwin,S.D. Thesis. University of Manchester. 1979.}
\ref{Unwin2}{Unwin,S.D. \jpa{13}{1980}{313}.}
\ref{DandB}{Dowker,J.S. and Banach,R. \jpa{11}{1979}{}.}
\ref{Obhukov}{Obhukov,Yu.N. \pl{109B}{1982}{195}.}
\ref{Kennedy}{Kennedy,G. \prD{23}{1981}{2884}.}
\ref{CandT}{Copeland,E. and Toms,D.J. \np {255}{1985}{201}.}
\ref{ELV}{Elizalde,E., Lygren, M. and Vassilevich, D.V. \jmp {37}{1996}{3105}.}
\ref{Malurkar}{Malurkar,S.L. {\it J.Ind.Math.Soc} {\bf16} (1925/26) 130.}
\ref{Glaisher}{Glaisher,J.W.L. {\it Messenger of Math.} {\bf18} (1889) 1.}
\ref{Anderson}{Anderson,A. \prD{37}{1988}{536}.}
 \ref{CandA}{Cappelli,A. and D'Appollonio,\pl{487B}{2000}{87}.}
 \ref{Wot}{Wotzasek,C. \jpa{23}{1990}{1627}.}
 \ref{RandT}{Ravndal,F. and Tollesen,D. \prD{40}{1989}{4191}.}
 \ref{SandT}{Santos,F.C. and Tort,A.C. \pl{482B}{2000}{323}.}
 \ref{FandO}{Fukushima,K. and Ohta,K. {\it Physica} {\bf A299} (2001) 455.}
 \ref{GandP}{Gibbons,G.W. and Perry,M. \prs{358}{1978}{467}.}
 \ref{Dow4}{Dowker,J.S. {\it Zero modes, entropy bounds and partition
functions.} hep-th\break /0203026.}
  \ref{Rad}{Rademacher,H. {\it Topics in analytic number theory,} 
Springer-Verlag,  Berlin,1973.}
  \ref{Halphen}{Halphen,G.-H. {\it Trait\'e des Fonctions Elliptiques}, Vol 1,
Gauthier-Villars, Paris, 1886.}
  \ref{CandW}{Cahn,R.S. and Wolf,J.A. {\it Comm.Mat.Helv.} {\bf 51} (1976) 1.}
  \ref{Berndt}{Berndt,B.C. \rmjm{7}{1977}{147}.}
  \ref{Hurwitz}{Hurwitz,A. \ma{18}{1881}{528}.}
  \ref{Hurwitz2}{Hurwitz,A. {\it Mathematische Werke} Vol.I. Basel, 
  Birkhauser, 1932.}
  \ref{Berndt2}{Berndt,B.C. \jram{303/304}{1978}{332}.}
  \ref{RandA}{Rao,M.B. and Ayyar,M.V. \jims{15}{1923/24}{150}.}
  \ref{Hardy}{Hardy,G.H. \jlms{3}{1928}{238}.}
  \ref{TandM}{Tannery,J. and Molk,J. {\it Fonctions Elliptiques},
   Gauthier-Villars, Paris, 1893--1902.}
  \ref{schwarz}{Schwarz,H.-A. {\it Formeln und Lehrs\"atzen zum Gebrauche..},
  Springer 1893.(The first edition was 1885.) The French translation by
Henri Pad\'e is
{\it Formules et Propositions pour L'Emploi...}, Gauthier-Villars, Paris,
1894}
  \ref{Hancock}{Hancock,H. {\it Theory of elliptic functions.} Vol I.
   Wiley, New York 1910.}
  \ref{watson}{Watson,G.N. \jlms{3}{1928}{216}.}
  \ref{MandO}{Magnus,W. and Oberhettinger,F. {\it Formeln und S\"atze},
  Springer-Verlag, Berlin 1948.}
  \ref{Klein}{Klein,F. {\it }.}
  \ref{AandL}{Appell,P. and Lacour,E. {\it Fonctions Elliptiques},
  Gauthier-Villars,
  Paris, 1897.}
  \ref{HandC}{Hurwitz,A. and Courant,C. {\it Allgemeine Funktionentheorie},
  Springer,
  Berlin, 1922.}
  \ref{WandW}{Whittaker,E.T. and Watson,G.N. {\it Modern analysis},
  Cambridge 1927.}
  \ref{SandC}{Selberg,A. and Chowla,S. \jram{227}{1967}{86}. }
  \ref{zucker}{Zucker,I.J. {\it Math.Proc.Camb.Phil.Soc} {\bf 82 }(1977) 111.}
  \ref{glasser}{Glasser,M.L. {\it Maths.of Comp.} {\bf 25} (1971) 533.}
  \ref{GandW}{Glasser, M.L. and Wood,V.E. {\it Maths of Comp.} {\bf 25} (1971)
  535.}
  \ref{greenhill}{Greenhill,A,G. {\it The Applications of Elliptic
  Functions}, MacMillan, London, 1892.}
  \ref{Weierstrass}{Weierstrass,K. {\it J.f.Mathematik (Crelle)}
{\bf 52} (1856) 346.}
  \ref{Weierstrass2}{Weierstrass,K. {\it Mathematische Werke} Vol.I,p.1,
  Mayer u. M\"uller, Berlin, 1894.}
  \ref{Fricke}{Fricke,R. {\it Die Elliptische Funktionen und Ihre Anwendungen},
    Teubner, Leipzig. 1915, 1922.}
  \ref{Konig}{K\"onigsberger,L. {\it Vorlesungen \"uber die Theorie der
 Elliptischen Funktionen},  \break Teubner, Leipzig, 1874.}
  \ref{Milne}{Milne,S.C. {\it The Ramanujan Journal} {\bf 6} (2002) 7-149.}
  \ref{Schlomilch}{Schl\"omilch,O. {\it Ber. Verh. K. Sachs. Gesell. Wiss.
  Leipzig}  {\bf 29} (1877) 101-105; {\it Compendium der h\"oheren Analysis},
  Bd.II, 3rd Edn, Vieweg, Brunswick, 1878.}
  \ref{BandB}{Briot,C. and Bouquet,C. {\it Th\`eorie des Fonctions Elliptiques},
  Gauthier-Villars, Paris, 1875.}
  \ref{Dumont}{Dumont,D. \aim {41}{1981}{1}.}
  \ref{Andre}{Andr\'e,D. {\it Ann.\'Ecole Normale Superior} {\bf 6} (1877) 265;
  {\it J.Math.Pures et Appl.} {\bf 5} (1878) 31.}
  \ref{Raman}{Ramanujan,S. {\it Trans.Camb.Phil.Soc.} {\bf 22} (1916) 159;
 {\it Collected Papers}, Cambridge, 1927}
  \ref{Weber}{Weber,H.M. {\it Lehrbuch der Algebra} Bd.III, Vieweg, 
  Brunswick 190  3.}
  \ref{Weber2}{Weber,H.M. {\it Elliptische Funktionen und algebraische Zahlen},
  Vieweg, Brunswick 1891.}
  \ref{ZandR}{Zucker,I.J. and Robertson,M.M.
  {\it Math.Proc.Camb.Phil.Soc} {\bf 95 }(1984) 5.}
  \ref{JandZ1}{Joyce,G.S. and Zucker,I.J.
  {\it Math.Proc.Camb.Phil.Soc} {\bf 109 }(1991) 257.}
  \ref{JandZ2}{Zucker,I.J. and Joyce.G.S.
  {\it Math.Proc.Camb.Phil.Soc} {\bf 131 }(2001) 309.}
  \ref{zucker2}{Zucker,I.J. {\it Siam J.Math.Anal.} {\bf 10} (1979) 192,}
  \ref{BandZ}{Borwein,J.M. and Zucker,I.J. {\it IMA J.Math.Anal.} {\bf 12}
  (1992) 519.}
  \ref{Cox}{Cox,D.A. {\it Primes of the form $x^2+n\,y^2$}, Wiley, New York,
  1989.}
  \ref{BandCh}{Berndt,B.C. and Chan,H.H. {\it Mathematika} {\bf42} (1995) 278.}
  \ref{EandT}{Elizalde,R. and Tort.hep-th/}
  \ref{KandS}{Kiyek,K. and Schmidt,H. {\it Arch.Math.} {\bf 18} (1967) 438.}
  \ref{Oshima}{Oshima,K. \prD{46}{1992}{4765}.}
  \ref{greenhill2}{Greenhill,A.G. \plms{19} {1888} {301}.}
  \ref{Russell}{Russell,R. \plms{19} {1888} {91}.}
  \ref{BandB}{Borwein,J.M. and Borwein,P.B. {\it Pi and the AGM}, Wiley,
  New York, 1998.}
 \end{putreferences}
\bye